\RequirePackage{fix-cm}
\documentclass[twocolumn]{svjour3}          
\smartqed  

\usepackage[utf8]{inputenc}
\usepackage{url}
\usepackage{graphicx}
\usepackage{balance}
\usepackage{color}
\usepackage{times,amsmath,epsfig}
\usepackage{amssymb}
\usepackage{textcomp}
\usepackage[linesnumbered,algoruled,boxed,lined]{algorithm2e}
\usepackage{cite}
\usepackage{multirow}
\usepackage{caption, subcaption}
\newcommand{\fanis}[1]{{\color{black}#1}}
\newcommand{\karima}[1]{{\color{black}#1}}
\begin{document}
\title{ProS: Data Series Progressive $k$-NN Similarity Search and Classification\\ with Probabilistic Quality Guarantees}

\institute{}

\author{Karima Echihabi         \and
		Theophanis Tsandilas    \and
		Anna Gogolou			\and
		Anastasia Bezerianos	\and
		Themis Palpanas
}

\institute{K. Echihabi \at
	Mohammed VI Polytechnic University (UM6P)\\
	\email{karima.echihabi@um6p.ma} 
	\and
	T. Tsandilas \at
	Université Paris-Saclay, CNRS, Inria, LISN\\
	\email{theophanis.tsandilas@lisn.upsaclay.fr}
	\and
	A. Gogolou \at
	Université Paris-Saclay, CNRS, Inria, LISN\\
	\email{a.gogolou@gmail.com}
	\and
	A. Bezerianos \at
	Université Paris-Saclay, CNRS, Inria, LISN\\
	\email{anastasia.bezerianos@lisn.upsaclay.fr}
\and
	T. Palpanas \at
	LIPADE, University of Paris \& French Univ. Institute (IUF)\\
	\email{themis@mi.parisdescartes.fr}	
}

\date{Received: date / Accepted: date}

\maketitle

\begin{abstract}
Existing systems dealing with the increasing volume of data series cannot guarantee interactive response times, even for fundamental tasks such as similarity search.
Therefore, it is necessary to develop analytic approaches that support exploration and decision making by providing progressive results, before the final and exact ones have been computed. Prior works lack both efficiency and accuracy when applied to large-scale data series collections. 
We present and experimentally evaluate ProS, a new probabilistic learning-based method that provides quality guarantees for progressive Nearest Neighbor (NN) query answering. 
We develop our method for $k$-NN queries and demonstrate how it can be applied with the two most popular distance measures, namely, Euclidean and Dynamic Time Warping (DTW).
We provide both initial and progressive estimates of the final answer that are getting better during the similarity search, as well suitable stopping criteria for the progressive queries. 
Moreover, we describe how this method can be used in order to develop a progressive algorithm for data series classification (based on a $k$-NN classifier), and we additionally propose a method designed specifically for the classification task.
Experiments with several and diverse synthetic and real datasets demonstrate that our prediction methods constitute the first practical solutions to the problem, significantly outperforming competing approaches. This paper was published in the VLDB Journal (2022). \end{abstract}

%
%




\section{Introduction}
\label{sec:intro}

\noindent
\textbf{Data Series.} 
Data series are ordered sequences of values measured and recorded from a wide range of human activities and natural processes~\cite{Palpanas2019}, such as seismic activity, 
or electroencephalography (EEG) signal recordings. 
The analysis of such sequences\footnote{If the dimension that imposes the ordering of the sequence is time then we talk about \emph{time series}. Though, a series can also be defined over other measures (angle in radial profiles, 
frequency in infrared spectroscopy, 
etc.). We use the terms \emph{time series}, \emph{data series}, and \emph{sequence} interchangeably.} is becoming increasingly challenging as their sizes often grow to multiple terabytes~\cite{DBLP:journals/sigmod/Palpanas15,DBLP:journals/dagstuhl-reports/BagnallCPZ19,DBLP:conf/edbt/EchihabiZP21,DBLP:journals/pvldb/EchihabiPZ21}. 

Data series analysis involves pattern matching~\cite{DBLP:journals/pvldb/ZoumpatianosIP15,journal/vldb/linardi19,eenergy21}, anomaly detection~\cite{journal/csur/Chandola2009,DBLP:journals/kais/YankovKR08,journal/vldb/Dallachiesa2014,DBLP:journals/datamine/LinardiZPK20,DBLP:conf/edbt/Gao0B20,norma,normajournal,series2graph,DBLP:conf/edbt/Gao0B20,DBLP:journals/csur/Blazquez-Garcia21,distrs2g,sand,DBLP:conf/kdd/Lu00ZK22,DBLP:journals/pvldb/PaparrizosKBTPF22,VUS,theseus}, frequent pattern mining~\cite{Rakthanmanon:2012:SMT,DBLP:journals/kais/GaoL19,VALMOD}, clustering~\cite{conf/kdd/Keogh1998,conf/sdm/Rodrigues2006,conf/icdm/Keogh2011,journal/pattrecog/Warren2005,DBLP:journals/tods/PaparrizosG17,DBLP:journals/datamine/LiLZ21}, and classification~\cite{journal/jmlr/Chen2009,DBLP:journals/datamine/SchaferL20,DBLP:journals/datamine/BagnallLBLK17,Yeh:2018:TSJ,DBLP:journals/datamine/LucasSPOZGPW19,DBLP:conf/sigmod/BoniolMRP22,iedeal}. 
Several algorithms relevant to these tasks rely on \emph{data series similarity}. 
The data-mining community has proposed several techniques, 
including many similarity measures (or distance measure algorithms), for calculating the distance between two data series~\cite{Ding:2008,DBLP:conf/ssdbm/MirylenkaDP17,DBLP:conf/sigmod/PaparrizosLEF20}, as well as corresponding indexing techniques and algorithms~\cite{Echihabi:2018,evolutionofanindex,DBLP:conf/edbt/EchihabiZP21}, in order to address scalability challenges.

\vspace{5pt} 
\noindent
\textbf{Data Series Similarity.} 
We observe that data series similarity is often domain- and visualization-dependent~\cite{Batista:2014:CEC,Gogolou:2019}, and in many situations, analysts depend on time-consuming manual analysis processes.
For example, neuroscientists manually inspect the EEG data of their patients, using visual analysis tools, so as to identify patterns of interest~\cite{Jing:2016,Gogolou:2019}.
In such cases, it is important to have techniques that operate within interactive response times~\cite{resp-times}, in order to enable analysts to complete their tasks easily and quickly.

In the past years, several visual analysis tools have combined visualizations with advanced data management and analytics techniques (e.g.,~\cite{Rahman:2017:ISE,DBLP:journals/pvldb/Kraska18}), albeit not targeted to data series similarity search. 
Moreover, we note that even though the data series management community is focusing on scalability issues, 
the state-of-the-art indexes currently used for scalable data series processing~\cite{Wang:2013:DDS,journal/kais/Camerra2014,Zoumpatianos:2016,journal/pvldb/kondylakis18,journal/vldb/linardi19} are still far from achieving interactive response times~\cite{Echihabi:2018,conf/vldb/echihabi2019}.

\begin{figure}[tb]
\centering
\footnotesize
\includegraphics[width=0.94\linewidth]{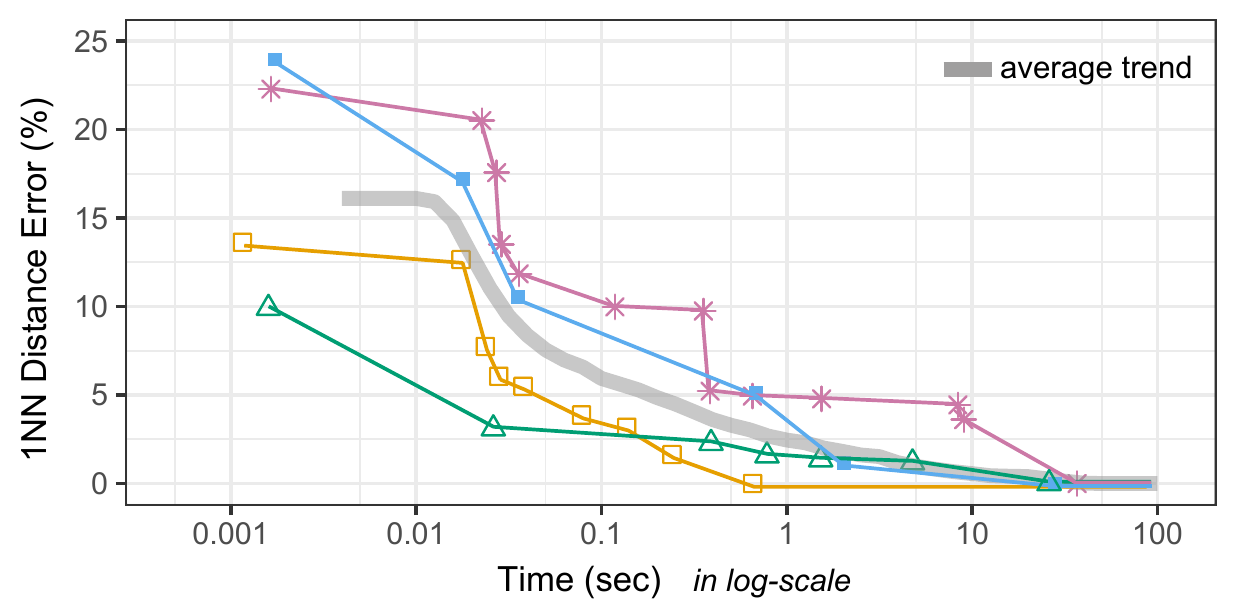}
\caption{Progression of $1$-NN distance error (Euclidean dist.) for 4 example queries (seismic dataset), using iSAX2+~\cite{journal/kais/Camerra2014}. The points in each curve represent approximate (intermediate points) or exact answers (last point) given by the algorithm. Lines end when similarity search ends. Thick gray line represents average trend over a random sample of 100 queries.}
\label{fig:motivation}
\vspace*{-0.4cm}
\end{figure}

\vspace{5pt} 
\noindent
\textbf{Progressive Results.} 
To allow for interactive response times when users analyze large data series collections, we need to consider progressive and iterative visual analytics approaches \cite{Badam:2017,Zgraggen:2017,Turkay:2017,DBLP:conf/edbt/GogolouTPB19}. 
Such approaches provide progressive answers to users' requests~\cite{Fisher:2012,Stolper:2014,Moritz:2017}, sometimes based on algorithms that return quick approximate answers~\cite{Ding:2016,Fekete:2016}. 
Their goal is to support exploration and decision making by providing progressive 
\karima{results. A progressive result is an intermediate answer that iteratively converges to the final, correct solution.}

Most of the above techniques consider approximations of aggregate queries on relational databases, with the exception of Ciaccia et al.~\cite{Ciaccia:1999,Ciaccia:2000}, who provide a probabilistic method for assessing how far an approximate answer is from the exact answer. 
Nevertheless, these works do not consider data series that are high-dimensional\footnote{\karima{The dimensionality of a data series is the length, or number of points in the series~\cite{Echihabi:2018}. In our context, by high-dimensional, we refer to series with dimensionality in the order of hundreds-thousands.}}. 
We note that the framework of Ciaccia et al.~\cite{Ciaccia:1999,Ciaccia:2000} does not explicitly target progressive similarity search. 
\karima{Furthermore, the approach has only been tested on datasets with up to 275K vectors with dimensionality up to 100, while we are targeting data series vectors in the order of hundreds of millions (in our experiments we provide results with up to 267M series), and with dimensionality that can exceed 1000 (in our experiments we provide results with up to 1280)}. 
Our experiments show that the probabilistic estimates that their methods~\cite{Ciaccia:1999,Ciaccia:2000} provide are inaccurate and cannot support progressive similarity search.

In this study, we demonstrate the importance of providing progressive 
similarity search results on large time series collections. 
Our results show that there is a gap between the time the 1st Nearest Neighbour (1-NN) is found and the time when the search algorithm terminates. 
In other words, users often wait without any improvement in their answers. We further show that high-quality approximate answers are found very early, e.g., in less than one second, so they can support highly interactive visual analysis tasks. 

Figure~\ref{fig:motivation} presents the approximate results of the iSAX2+ index~\cite{journal/kais/Camerra2014} for four example queries on a 100M data series collection with seismic data~\cite{url/data/seismic}, where we show the evolution of the approximation error as a percentage of the exact $1$-NN distance.
We observe that the algorithm provides approximate answers within a few milliseconds, and those answers gradually converge to the exact answer, which is the distance of the query from the $1$-NN. 
Interestingly, the $1$-NN is often found in less than 1 sec (e.g., see yellow line), but it takes the search algorithm much longer to verify that there is no better answer and terminate. This finding is consistent with previously reported results~\cite{Ciaccia:2000,DBLP:conf/edbt/GogolouTPB19}. 

Several similarity-search algorithms, such as the iSAX2+ index~\cite{journal/kais/Camerra2014} and the DSTree~\cite{Wang:2013:DDS} (the two top performers in terms of data series similarity search~\cite{Echihabi:2018}), provide very quick approximate answers. In this paper, we argue that such algorithms can be used as the basis for supporting progressive similarity search.  Unfortunately, these algorithms do not provide any guarantees about the quality of their approximate answers, while our goal is to provide such guarantees.

\vspace{5pt} 
\noindent
\textbf{Proposed Approach.} 
We develop \emph{ProS}, the first progressive \fanis{approach} for sequence search and classification with probabilistic quality guarantees, which \fanis{is} scalable to very large data series collections.  
Our goal is to predict how much improvement is expected when the algorithms are still running. 
Communicating this information to users will allow them to terminate a progressive analysis task early and save time.

\begin{figure*}[tb]
	\centering
	\footnotesize
	\includegraphics[width=1.0\textwidth]{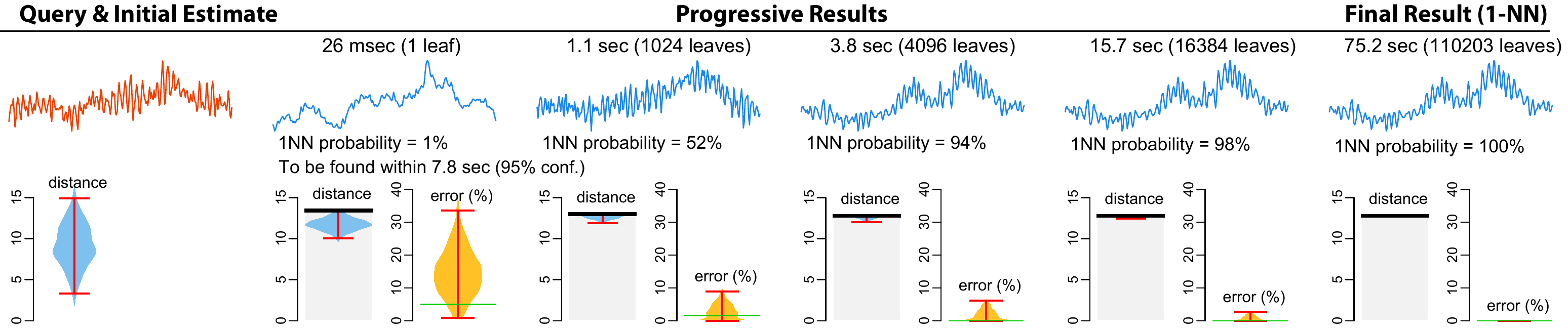}
	\caption{\fanis{A query example from the seismic dataset showing the evolution of estimates based on our approach. The thick black lines show the distance of the current approximate answer. The red error bars represent $95\%$ prediction intervals. The green line over the predicted distribution of distance errors shows the real error -- it is unknown during the search and is shown here for illustration purposes. Estimates are based on a training set of 100 queries, as well as 100 random witnesses for initial estimates. We use the iSAX2+ index.}}
	\label{fig:examples-vis}
\end{figure*}

\fanis{Figure~\ref{fig:examples-vis} showcases our approach with an example on real data. An analyst enters a seismic pattern as a query (in red) and immediately (response times reported at the top of the figure) receives progressive approximations of its 1-NN (in blue). 
In addition to these progressive answers, the system also provides estimates of the current distance error: the blue distributions~\cite{yarrr} estimate the absolute distance error; while the yellow distributions estimate the relative distance error. 
Observe that the initial distance estimate is rather uncertain, but estimates become precise at the early stages of the search. 
The system can further communicate a probability of whether the current answer is exact and predict when the exact answer is expected with a certain confidence level. 
In this example, the query terminates after 75.2sec, but the above predictions can give confidence to the user that the current answer is very close to the $1$-NN much earlier (i.e., almost one order of magnitude faster). 
The user can then decide to stop the query.}

The challenge is how to derive such predictions. If we further inspect Figure~\ref{fig:motivation}, we see that similarity search answers progressively improve, but improvements are not radical. 
The error of the first approximate answer (when compared to the final exact answer) is on average $16\%$, which implies that approximate answers are generally not very far from the $1$-NN. 
We show that this behavior is more general and can be observed across different datasets and different similarity search algorithms~\cite{Zoumpatianos:2016,Wang:2013:DDS}. 
We further show that the distance of approximate answers can help us predict the time that it takes to find the exact answer. 
Our approach 
\karima{describes} these behaviors through statistical models. 
We then use these models to estimate the error of a progressive answer, assess the probability of an early exact answer, and provide upper bounds for the time needed to find the $k$-NN. 
We also explore query-sensitive models that predict a probable range of the $k$-NN distance before the search algorithm starts, and then is progressively improved as new answers arrive. 
We further provide reliable stopping criteria 
for terminating searches with probabilistic guarantees about the distance error or the number of exact answers.

In addition to similarity search, we address the problem of $k$-NN classification. We show how the early termination of $k$-NN similarity search can be employed to  lead to time savings for $k$-NN classification. 
Moreover, we propose probabilistic guarantees for the exact class itself, as well, which allows us to achieve even larger savings. 

We note that earlier approaches~\cite{Ciaccia:1999,Ciaccia:2000} do not solve the problem, since they support bounds only for \emph{distance} errors, they do not update their estimates during the course of query answering, and they do not scale with the data size.

\vspace{5pt} 
\noindent
\textbf{Contributions.} 
Our key contributions
are as follows.

\begin{itemize}
\item 
We formulate the problem of \emph{progressive data series similarity search}, and provide definitions specific to the context of data series.

\item 
We investigate statistical methods, based on regression (linear, quantile, and logistic) and multivariate kernel density estimation, for supporting progressive similarity search based on a small number (50 - 200) of training queries. 
We show how to apply them to derive estimates for the $k$-NN distance (distance error), the time to find the $k$-NN, and the probability that the progressive $k$-NN answer is correct.

\item
We further develop stopping criteria that can be used to stop the search long before the normal query execution ends. 
These criteria make use of distance error estimates, probabilities of exact answers, and time bounds for exact answers.
We show how our criteria can be applied with the two most popular data series distance measures, namely, Euclidean and Dynamic Time Warping (DTW).

\item
Moreover, we describe how our approach extends to data series $k$-NN classification. 
In particular, we derive probabilistic guarantees and develop stopping criteria for the exact class of a progressive $k$-NN classification. 


\item
We perform an extensive experimental evaluation with several and diverse synthetic and real datasets. 
The results demonstrate that our solutions dominate the previous approaches, 
provide accurate probabilistic bounds, and lead to significant time improvements with well-behaved guarantees for errors. 
Source code and datasets are publicly available~\cite{webpage}.

\end{itemize}


\karima{
This paper extends our previous work~\cite{DBLP:conf/sigmod/GogolouTEBP20} in the following directions:

\begin{itemize}
	
	\item We extend the original method that was designed for 1-NN similarity queries to support k-NN queries. 

	\item In addition to the Euclidean distance measure, we now also study in detail the Dynamic Time Warping (DTW) distance in the context of progressive data series similarity search. 

	\item Apart from pattern matching using similarity search, we now propose methods for progressive classification, as well, which is a very popular analysis task for data series collections. 

	\item We expand the empirical evaluation of our methods by adding five new synthetic and real datasets, as well as several new experiments and discussions.
		
	\item We expand the discussions of the related work, which helps draw a more complete picture of the research area relevant to our work.  

\end{itemize}
}

\noindent
\textbf{Paper Structure.} 
The rest of this paper is organized as follows. 
Section~\ref{sec:related-work} summarizes related work on data series similarity search and progressive visual analytics, and Section~\ref{sec:def} presents background terminology. 
In Section~\ref{sec:progressive}, we define progressive similarly search and introduce the main concepts for supporting prediction with probabilistic guarantees. 
Then, in \fanis{Section~\ref{sec:methods}}, we describe our methods for estimating a $k$-NN distance before and during the execution of a similarity search query, and in Section~\ref{sec:knn-classification-approach} the corresponding methods for $k$-NN classification. 
Section~\ref{sec:experiments} presents an extensive evaluation of all proposed methods. 
Finally, we conclude in Section~\ref{sec:conclusion}, where we also propose directions for future work.


\vspace{-2pt} 
\section{Related Work}
\label{sec:related-work}



\textbf{Similarity Search.} 
%
Several measures have been proposed for computing similarity between data series~\cite{Ding:2008,DBLP:conf/ssdbm/MirylenkaDP17}. Among them, Euclidean Distance (ED)~\cite{conf/sigmod/Faloutsos1994}, which performs a point-by-point value comparison between two time series, is one of the most popular. ED can be combined with data normalization (often \emph{z-normalization}~\cite{Goldin:1995:SQT}), in order to consider as similar patterns that may vary in amplitude or value offset. 
In our work, we focus on ED because it is effective, and leads to efficient solutions for large datasets~\cite{Ding:2008,Echihabi:2018}. 
\fanis{We also extend our approach to DTW~\cite{DBLP:conf/kdd/RakthanmanonCMBWZZK12}, which is very popular in practice, and more suitable than ED for certain applications, especially in classification tasks~\cite{DBLP:journals/datamine/BagnallLBLK17}.}

 

The human-computer interaction community has focused on the interactive visual exploration and querying of data series. 
These querying approaches are visual, often on top of line chart visualizations~\cite{Tufte:1986:VDQ}, and rely either on the interactive selection of part of an existing series (e.g.,~\cite{Buono:2008:ISS}), or on sketching patterns to search for (e.g.,~\cite{Correll:2016,Mannino:2018:ETS}). 
This line of work is orthogonal to our approach, which considers approximate and progressive results from these queries when interactive search times are not possible. 

%
\vspace{5pt}
\noindent
\textbf{Optimized and Approximate Similarity Search.} 
The data-base community has optimized similarity search methods by using index structures~\cite{conf/sigmod/Faloutsos1994,Ciaccia:1998:CMS,Wang:2013:DDS,Camerra:2010:IIM,Zoumpatianos:2016,dpisaxjournal,journal/pvldb/kondylakis18,DBLP:journals/vldb/KondylakisDZP19,journal/vldb/linardi19,ulissejournal,DBLP:conf/gis/Chatzigeorgakidis19,DBLP:conf/edbt/Chatzigeorgakidis21,seanet} 
or fast sequential scans~\cite{Rakthanmanon:2012:SMT}. 
Recently, Echihabi et al. \cite{Echihabi:2018,journal/pvldb/echihabi2019} compared the efficiency of these methods under a unified experimental framework, showing that there is no single best method that outperforms all the rest.
In this study, we focus on the popular centralized solutions, though, our results naturally extend to parallel and distributed solutions~\cite{conf/bigdata/peng18,parisplus,messi,messijournal,sing,dpisax,dpisaxjournal,DBLP:journals/kais/LevchenkoKYAMPS21,hercules}, 
since these solutions are based on the same principles and mechanisms as their centralized counterparts.
Moreover, we focus on (progressive answers for) exact query answering. 
Given enough time, all answers we produce are exact, which is important for several applications~\cite{Palpanas2019}. 
In this context, progressive answers help to speed-up exact queries by stopping execution early, when it is highly probable that the current progressive answer is the exact one.
Note that several data series similarity search methods support approximate query answering that can produce increasingly more accurate answers as time goes by~\cite{Wang:2013:DDS,journal/kais/Camerra2014,Zoumpatianos:2016,DBLP:journals/vldb/KondylakisDZP19,ulissejournal,DBLP:journals/is/FerhatosmanogluTAA06}, though, none of them provides quality guarantees on the answers. 
In this work, we focus on the iSAX2+~\cite{journal/kais/Camerra2014} and DSTree~\cite{Wang:2013:DDS} methods, which exhibit superior performance at the similarity search task~\cite{Echihabi:2018,journal/pvldb/echihabi2019}.

In parallel to our work, Li et al.~\cite{msearlytermination} proposed a machine learning method, 
developed on top of inverted-file (IVF~\cite{DBLP:journals/pami/JegouDS11} and IMI~\cite{DBLP:journals/pami/BabenkoL15}) and k-NN graph (HNSW~\cite{DBLP:journals/pami/MalkovY20}) similarity search techniques, that solves the problem of early termination of approximate NN queries, while achieving a target recall. 
In contrast, our approach employs similarity search techniques based on data series indices~\cite{journal/pvldb/echihabi2019}, and with a very small training set (up to 200 training queries in our experiments), provides guarantees with per-query probabilistic bounds along different dimensions: on the distance error, on whether the current answer is the exact one, and on the time needed to find the exact answer.

\vspace{5pt}
\noindent
{\textbf{$k$-NN Classification.}}
Similarity-based classification (e.g., $k$-NN Classifier) is a supervised task consisting of assigning a label to a new item based on the majority vote of its neighbors among the set of labeled training samples. 
It is used in a variety of domains, such as bioinformatics for protein classification~\cite{conf/ismb/ankerst1999}, computer vision for object recognition~\cite{journal/jmlr/Chen2009}, text mining for web page categorization~\cite{conf/iral/kwon2000}, remote sensing~\cite{DBLP:conf/icdm/PetitjeanFWNCK14}, and social media for image classification~\cite{DBLP:journals/corr/abs-2106-09672}.
We note that, even though lots of work has been dedicated into developing data series classification algorithms, the $k$-NN classifier remains a strong baseline~\cite{DBLP:journals/datamine/BagnallLBLK17} and the only viable solution in use-cases with (limited hardware resources and) very large amounts of data~\cite{DBLP:conf/icdm/PetitjeanFWNCK14}.






To the best of our knowledge, the idea of progressive classification has not been carefully studied before. 
Previous work has looked at the problem of classifying images at multiple resolutions~\cite{DBLP:conf/icassp/CastelliLTK96}, but does not propose a progressive query answering framework, nor does it provide quality guarantees.

\vspace{5pt} 
\noindent
\textbf{Progressive Visual Analytics.}
Fekete and Primet~\cite{Fekete:2016} provide a summary of the features of a progressive system; three of them are particularly relevant to progressive data series search: 
(i) progressively improved answers; (ii) feedback about the computation state and costs; and (iii) guarantees of time and error bounds for progressive and final results.
%
Systems that support big data visual exploration include 
Pangloss~\cite{Moritz:2017} that provides quick approximate results of aggregation queries, Falcon~\cite{Moritz:2019:FBI} that prefetches data for brushing and linking actions, and IncVisage~\cite{Rahman:2017:ISE} that 
progressively reveals salient features in heatmap and trendline visualizations. 

Systems that provide progressive results are appreciated by users due to their quick feedback~\cite{Badam:2017,Zgraggen:2017}. Nevertheless, there are some caveats. Users can be mislead into believing false patterns~\cite{Moritz:2017,Turkay:2017} with early progressive results. It is thus important to communicate the progress of ongoing computations~\cite{Angelini2018ARA,Schulz:2016:EVP}, including the uncertainty and convergence of results~\cite{Angelini2018ARA} and guarantees on time and error bounds~\cite{Fekete:2016}.  
Previous work provides such uncertainty and guarantees in relational databases and aggregation type queries \cite{DBLP:conf/sigmod/HellersteinHW97,DBLP:journals/tods/JermaineAPD08,DBLP:series/isrl/WuOT13}.  

Closer to the context of data series, Ciaccia and Patella \cite{Ciaccia:2000} studied similarity search queries over general multi-dimensional spaces and proposed a probabilistic approach for computing the uncertainty of partial similarity search results. 
We discuss their approach in the following section.


\section{Preliminaries and Background}\label{sec:def}



A \textit{data series} $S(p_1,p_2,...,p_\ell)$ is an ordered sequence of 
\karima{real-valued points} with \fanis{length $\ell$.} 
A data series of length $\ell$ can also be represented as a single point in an $\ell$-dimensional space. 
For this reason, the values of a data series are often called \emph{dimensions}, and its length $\ell$ is called \emph{dimensionality}. 
%
We use $\mathbb{S}$ to denote a \textit{data series collection} (or \textit{dataset}). 
We refer to the size $n=|\mathbb{S}|$ of a data series collection as \textit{cardinality}. 
\karima{In this paper, we focus on datasets with a very large number of regularly sampled data series, with no uncertainty in the values~\cite{DBLP:conf/ssdbm/AssfalgKKR09,DBLP:conf/edbt/YehWYC09,DBLP:conf/kdd/SarangiM10,DBLP:journals/pvldb/DallachiesaNMP12,journal/vldb/Dallachiesa2014}, and no missing values~\cite{DBLP:conf/edbt/WellenzohnBDGM17,DBLP:journals/pvldb/BansalDS21}, which means that we do not need to encode the attribute describing the dimension of the sequence (e.g., the timestamps when the dimension is time).
While the techniques used in this paper are designed for series of equal length, our models could be extended to support series of variable length (e.g., following the ideas proposed by the ULISSE index~\cite{ulissejournal}).}

\vspace{5pt} 
\noindent{\bf Distance Measures.} A data series \textit{distance} $d(S_1,S_2)$ is a function that measures the dissimilarity of two data series $S_1$ and $S_2$, or alternatively, the dissimilarity of two data series subsequences. 
As mentioned in Sec~\ref{sec:related-work}, we chose Euclidean Distance (ED) as a measure due to its popularity and efficiency~\cite{Ding:2008}. 


\vspace{5pt} 
\noindent{\bf Similarity Search Queries.}
Given a dataset $\mathbb{S}$, a \textit{query} series $Q$, and a distance function $d(\cdot,\cdot)$, a \textit{k-Nearest-Neighbor (}$k$-NN\textit{) query} identifies the $k$ series in the dataset with the smallest distances to $Q$. The 1st Nearest Neighbor (1-NN) is the series in the dataset with the smallest distance to $Q$.


Similarity search can be \textit{exact}, when it produces answers that are always correct, or \textit{approximate}, when there is no such strict guarantee. 
A \textit{$\delta$-$\epsilon$-approximate algorithm} guarantees that its distance results will have a relative error no more than $\epsilon$ with a probability of at least $\delta$~\cite{Echihabi:2018}. 
We note that only a couple of approaches~\cite{Arya:1998,Ciaccia:2000} provide such guarantees. 
Yet, their accuracy has never been tested on the range of dimensions and dataset sizes that we examine here. 

\vspace{5pt} 
\noindent{\bf Similarity Search Methods.} 
Most data series similarity search techniques~\cite{conf/sigmod/Faloutsos1994,Ciaccia:1998:CMS,Wang:2013:DDS,Camerra:2010:IIM,Zoumpatianos:2016,dpisaxjournal,conf/bigdata/peng18,journal/pvldb/kondylakis18,journal/vldb/linardi19,evolutionofanindex} use an index, which enables scalability.
The index can offer quick approximate answers by traversing a single path of the index structure to visit the single most promising leaf, from where we select the \textit{best-so-far (bsf)} answer: this is the candidate answer in the leaf that best matches (has the smallest distance to) the query. 
The bsf may, or may not be the final, exact answer: in order to verify, we need to either prune, or visit all the other leaves of the index. 
Having a good first bsf (i.e., close to the exact answer) leads to efficient pruning. 

In the general case, approximate data series similarity search algorithms do not provide guarantees about the quality of their answers. 
In our work, we illustrate how we can efficiently 
provide such guarantees, with probabilistic bounds. 

We focus on index-based approaches that support both quick approximate, and slower but exact, similarity search results. 
In this work, we adapt the state-of-the-art data series indexes iSAX2+~\cite{journal/kais/Camerra2014} and DSTree~\cite{Wang:2013:DDS}, which have been shown to outperform the other data series methods in query answering~\cite{Echihabi:2018}, and we demonstrate that our techniques are applicable to both indexes. 
We provide below a succinct description of the iSAX2+ and DSTree approaches.

The iSAX2+~\cite{journal/kais/Camerra2014} index organizes the data in a tree structure, where the leaf nodes contain the raw data and each internal node summarizes the data series that belong to its subtree using a representation called \textit{Symbolic Aggregate Approximation (SAX)}~\cite{conf/dmkd/LinKLC03}. SAX transforms a data series using Piecewise Aggregate Approximation (PAA)~\cite{journal/kais/Keogh2001} into equi-length segments, where each segment is associated with the mean value of its points, then represents the mean values using a discrete set of symbols for a smaller footprint. 

DSTree~\cite{Wang:2013:DDS} is also a tree-based index that stores raw data in the leaves and summaries in internal nodes. 
Contrary to iSAX2+, DSTree does not support bulkloading, intertwines data segmentation with indexing and uses \textit{Extended Adaptive Piecewise Approximation (EAPCA)}~\cite{Wang:2013:DDS} instead of SAX. With EAPCA, a data series is segmented using 
APCA~\cite{journal/acds/Chakrabarti2002} into varying-length segments, then each segment is represented with its mean and standard deviation values.

Since the query answering time depends on the data distribution~\cite{conf/kdd/Zoumpatianos2015}, and both iSAX2+ and DSTree can produce unbalanced index trees, we provide below an index-invariant asymptotic analysis on the lower/upper bounds of the query runtime. 
As we consider large on-disk datasets, the query runtime is I/O bound; thus we express complexity in terms of I/O~\cite{analysis-kanellakis,analysis-hellerstein}, using the dataset size \fanis{$n$}, the index leaf threshold $th$ and the disk block size $B$. 
Consider an index over a dataset of size \fanis{$n$} such that each index leaf contains at most $th$ series \fanis{($th \ll n$)}. 
We count one disk page access of size $B$ as one I/O operation (for simplicity, we use $B$ to denote the number of series that fit in one disk page). 
Note that both the iSAX2+ and DSTree indexes fit the entire index tree in-memory; the leaves point to the raw data on-disk.

\underline{Best Case}. The best case scenario occurs when one of the children of the index root is a leaf, containing one data series. 
In this case, the approximate search will incur $\Theta(1)$ I/O operation. 
In the best case, exact search will prune all other nodes of the index and thus will also incur $\Theta(1)$ disk access. 

\underline{Worst Case}. Approximate search always visits one leaf. 
Therefore, the worst case occurs when the leaf is the largest possible, i.e., it contains $th$ series, in which case approximate search incurs $\Theta(th/B)$ I/O operations. 
For exact search, the worst case occurs when the algorithm needs to visit every single leaf of the index. 
This can happen when the index tree has \fanis{$n-th+1$} leaves (i.e., each leaf contains only one series, except for one leaf with $th$ series), as a result of each new series insertion causing a leaf split where only one series ends up in one of the children. 
Therefore, the exact search algorithm will access all the leaves, 
and will incur $\Theta(N)$ I/O operations. 
(Note that this is a pathological case that would happen when all series are almost identical: 
in this case, indexing and similarity search are not useful anyways.)

\vspace{5pt} 
\noindent{\bf Baseline Approach.}
We briefly describe here the probabilistic approach of Ciaccia et al.~\cite{Ciaccia:1998:CMS,Ciaccia:1999,Ciaccia:2000}. 
%
Based on Ciaccia et al.~\cite{Ciaccia:1998:CMS}, a dataset $\mathbb{S}$ (a data series collection in our case) can be seen as a random sample drawn from a large population $\mathcal{U}$ of points in a high-dimensional space. 
Being a random sample, a large dataset is expected to be representative of the original population. 
Given a query $Q$, let $f_Q(x)$ be the probability density function that gives the relative likelihood that $Q$'s distance from a random series drawn from $\mathcal{U}$ is equal to $x$. 
Likewise, let $F_Q(\cdot)$ be its cumulative probability function. 
Based on $F_Q(\cdot)$, 
we can derive the cumulative probability function $G_{Q,n}(\cdot)$ for $Q$'s $k$-NN distances in a dataset of size $n=|\mathbb{S}|$. 
For $1$-NN similarity search, we have:
\begin{equation}\label{eq:G_Q}
\small
G_{Q,n}(x) = 1 - (1 - F_Q(x))^n
\end{equation}
We now have a way to construct estimates for $1$-NN distances. 
%
Unfortunately, $f_Q(\cdot)$, and thus $F_Q(\cdot)$, are not known.
The challenge is how to approximate them from a given dataset. 
We discuss two approximation methods:

\vspace{5pt} 
\noindent{1. \emph{Query-Agnostic Approximation.}} 
For high-dimensional spaces, a large enough sample from the overall distribution $f(\cdot)$ of pairwise distances in a dataset provides a reasonable approximation for $f_Q(\cdot)$~\cite{Ciaccia:1998:CMS}. 
This approximation can then be used to evaluate probabilistic stopping-conditions by taking sampling sizes between $10\%$ and $1\%$ (for larger datasets)~\cite{Ciaccia:2000}. 

\vspace{5pt} 
\noindent{2. \emph{Query-Sensitive Approximation.}} 
The previous method does not take the query into account. 
A  query-sensitive approach is based on a training set of reference queries, called \textit{witnesses}. 
Witnesses can be randomly drawn from the dataset, or selected with the GNAT algorithm~\cite{Brin:1995}, which 
identifies the $n_w$ points that best cover a multidimensional (metric) space based on an initial random sample of $3 n_w$ points.
Given that close objects have similar distance distributions, Ciaccia et al.~\cite{Ciaccia:1999} approximate $f_Q(\cdot)$ by using a weighted average of the distance distributions of all the witnesses. 


\vspace{5pt} 
The above methods have major limitations. First, since their $1$-NN distance estimates are static, they are less appropriate for progressive similarity search. 
Second, a good approximation of $F_Q(\cdot)$ does not necessarily lead to a good approximation of $G_{Q,n}(\cdot)$. 
This is especially true for large datasets, as the exponent term $n$ in Equation~\ref{eq:G_Q} will inflate even tiny approximation errors. 
Note that $G_{Q,n}(\cdot)$ can be thought of as a scaled version of $F_Q(\cdot)$ that zooms in on the range of the lowest distance values. 
If this narrow range of distances is not accurately approximated, the approximation of $G_{Q,n}(\cdot)$ will also fail. Our own evaluation demonstrates this problem.
Third, they require the calculation of a large number of distances. Since the approximation of $G_{Q,n}(\cdot)$ is sensitive to errors in large datasets (see above), a rather large number of samples is required in order to capture the frequency of the very small distances. 

\begin{table}[tb]
\centering
\caption{Table of symbols}
\label{SymbolTable}
\scalebox{0.83}{
\begin{tabular}{r|l}
{\bf Symbol} & {\bf Description} \\
\hline
$S$, $Q$					& data series, query series \\
$\ell$						& length of a data series \\
$\mathbb{S}$				& data series collection (or dataset)\\
$n=|\mathbb{S}|$			& number of series in $\mathbb{S}$ \\
$R(t)$						& progressive answer at time $t$ \\
$c_Q(t)$ 					& class of $k$-NN classification at time $t$\\
\emph{k-NN}, $knn(Q)$		& $k^{th}$ Nearest Neighbor of $Q$ \\
$d_{Q,R}(t)$, $d(Q,R(t))$		& distance between $Q$ and $R(t)$\\
$d_{Q,knn}$, $d(Q,knn(Q))$				& distance between $Q$ and its $k$-NN\\
${\epsilon}_Q(t)$			& relative distance error of $R(t)$ from $k$-NN \\
$\epsilon^f_{Q}(t)$			& relative family-wise distance error\\
$p_{Q}(t)$				& probability that  $R(t)$ is exact (i.e., the $k$-NN)\\
$p_{c_Q}(t)$				& probability that the class $c_Q(t)$ is exact\\	 
$t_{Q}$					& time to find the $k$-NN\\
$\tau_{Q,\phi}$				& time to find the $k$-NN with probability $1 - \phi$\\
$\tau_{Q,\theta,\epsilon}$ 		& time for which ${\epsilon}_Q(t) < \epsilon$ with confidence $1-\theta$\\
$\hat{\bullet}$				& estimate of $\bullet$ \\
$I_Q(t)$					& information at time $t$ \\
$h_{Q,t}(x)$				& probability density function of Q's distance \\
							& from its $k$-NN, given information $I_Q(t)$\\
$H_{Q,t}(x)$				& cumulative distribution function of Q's\\
							& distance from its $k$-NN, given $I_Q(t)$\\
$f_Q(x)$					& probability density function of $Q$'s distance \\
							& from a random series in $\mathbb{S}$ \\
$F_Q(x)$					& cumulative distribution function of $Q$'s\\
							&  distance from a random series in $\mathbb{S}$\\ 
$G_{Q,n}(x)$				& cumulative distribution function of $Q$'s\\
							& distance from its $k$-NN\\
$\mathcal{W}$				& set of witness series \\
$n_w=|\mathcal{W}|$			& number of witnesses in $|\mathcal{W}|$ 
\end{tabular}
}
\end{table}

\noindent{\bf k-NN Classification.} Given a training dataset $\mathbb{S}$ with individual data series allocated to a class in $C = \{c_1, c_2..., c_L\}$ and a new data series $Q$, a $k$-NN classifier assigns to $Q$ the most common class $c_Q \in C$ among its $k$ nearest neighbors in the training dataset. 
As a consequence, $k$-NN classification fully relies on $k$-NN similarity search, and therefore, there exists a direct link with all the methods that we describe below.


\section{Progressive Similarity Search}
\label{sec:progressive}

We define progressive similarity search for $k$-NN queries\footnote{We define the problem using $k$-NN, but for simplicity use $k=1$ in the rest of this paper. We defer the discussion of the general case to future work.}.
(Table~\ref{SymbolTable} summarizes the symbols we use in this paper.)
\begin{definition} \label{def:progressive}
Given a $k$-NN query $Q$,  a data series collection $\mathbb{S}$, and a time \textit{quantum} $q$, a \textit{\textbf{progressive similarity-search algorithm}} produces results $R(t_1), R(t_2), ..., R(t_z)$ at time points $t_1, t_2, ..., t_z$, where $t_{i+1} - t_i \le q$, such that\\ $d(Q, R(t_{i+1})) \le d(Q, R(t_{i}))$. 
\end{definition}
We borrow the quantum $q$ parameter from Fekete and Primet \cite{Fekete:2016}. 
It is a user-defined parameter that determines how frequently users require updates about the progress of their search. 
Although there is no guarantee that distance results will improve at every step of the progressive algorithm, the above definition states that a progressive distance will \emph{never} deteriorate. 
This is an important difference of progressive similarity search compared to other progressive computation mechanisms, where results may fluctuate before they eventually converge, which may lead users to making wrong decisions based on intermediate results~\cite{Fekete:2016,DBLP:conf/sigmod/ChaudhuriDK17,DBLP:conf/sigmod/GuoBK17}. 

Clearly, progressive similarity search can be based on approximate similarity search algorithms -- a progressive result is simply an approximate (best-so-far) answer that is updated over time.
A progressive similarity search algorithm is also exact if the following condition holds:
\begin{equation}
\small
\lim_{t\to\infty} d(Q, R(t)) = d(Q, knn(Q))
\end{equation}
\noindent where $knn(Q)$ represents the $k$-NN of the query series $Q$. 

According to the above condition, the progressive algorithm will always find an exact answer. However, there are generally no strong guarantees about how long this can take. Ideally, a progressive similarity search algorithm will find good answers very fast, e.g., within interactive times, and will also converge to the exact answer without long delays. Even so, in the absence of information, users may not be able to trust a progressive result, no matter how close it is to the exact answer. 

In this paper, we investigate exactly this problem. Specifically, we seek to provide guarantees about: (i) How close is the progressive answer to the exact answer? (ii) What is the probability that the current progressive answer is the exact answer? (iii) When is the search algorithm expected to find the exact answer? 


\subsection{Progressive Distance Estimates}
Given a progressive answer $R(t)$ to a $k$-NN query at time $t$, we are interested in knowing how far from the $k$-NN this answer is. For simplicity, we will denote the $k$-NN distance to the query as $d_{Q,knn}\equiv d(Q, knn(Q))$ and the distance between $R(t)$ and the query as $d_{Q,R}(t) \equiv d(Q, R(t))$. Then, the relative distance error \fanis{is
$\epsilon_Q(t) = \frac{d_{Q,R}(t)}{d_{Q,knn}(t)} - 1$.}
Given that this error is not known, our goal is to find an estimate $\hat{\epsilon}_Q(t)$. However, finding an estimate for the relative error is not any simpler than finding an estimate $\hat{d}_{Q,knn}(t)$ of the actual $k$-NN distance. We concentrate on this latter quantity for our analysis below. Though, since $d_{Q,R}(t)$ is known, deriving the distance error estimate $\hat{\epsilon}_Q(t)$ from the $k$-NN distance estimate $\hat{d}_{Q,knn}(t)$ is straightforward:
\begin{equation}\label{eq:error}
\small
 \hat{\epsilon}_Q(t) = \frac{d_{Q,R}(t)}{\hat{d}_{Q,knn}(t)} - 1
\end{equation}

We represent progressive similarity-search estimates as probability distribution functions.
\begin{definition} \label{def:progressive-distribution}
Given a $k$-NN query $Q$,  a data series collection $\mathbb{S}$, and a progressive similarity-search algorithm, a \textit{\textbf{progressive $k$-NN distance estimate}} $\hat{d}_{Q,knn}(t)$ of the $k$-NN distance at time $t$ is a probability density function:
\begin{equation}\label{eq:density}
\small
 h_{Q,t}(x) = Pr\{ d_{Q,knn} = x\ |\ I_Q(t)\} 
\end{equation}
\noindent This equation gives the conditional probability that $d_{Q,knn}$ is equal to $x$, given information $I_Q(t)$.
\end{definition}

We expect that progressive estimates will converge to $d_{Q,knn}$ (i.e., $\hat{\epsilon}_Q(t)$ will converge to zero).
Evidently, the quality of an estimate at time $t$ depends on the information $I_Q(t)$ that is available at this moment. 
In Section~\ref{sec:methods}, we investigate different types of information we can use for this. 

Given the probability density function in Equation~\ref{eq:density}, we can derive a point estimate that gives the \emph{expected} $k$-NN distance, or an interval estimate in the form of a \emph{prediction interval} (PI). 
Like a confidence interval, a prediction interval is associated with a confidence level. 
Given a confidence level $1 - \theta$, we expect that approximately $(1 - \theta) \times 100 \%$ of the prediction intervals we construct will include the true $k$-NN distance. 
Note that although a confidence level can be informally assumed as a probability (i.e., what is the likelihood that the interval contains the true $k$-NN distance?), this assumption may or may not be strictly correct. 
Our experiments evaluate the frequentist behavior of such intervals.  
   
To construct a prediction interval with confidence level $1-\theta$ over a density distribution function $h_{Q,t}(\cdot)$, we derive the cumulative distribution function:
\begin{equation}\label{eq:cumulative}
\small
H_{Q,t}(x) = Pr\{ d_{Q,knn} \le x\ |\ I_Q(t)\}
\end{equation}
\noindent From this, we take the $\theta/2$ and $(1 - \theta/2)$ quantiles that define the limits of the interval.

\subsection{Guarantees for Exact Answers}
\label{sec:exact}
Users may also need guarantees about the exact $k$-NN. We investigate two types of probabilistic guarantees for exact answers. First, at any moment $t$ of the progressive search, we assess the probability $p_{Q}(t)$ that the exact answer has been found, given information $I_Q(t)$: 
\begin{equation}\label{eq:probability}
\small
 p_{Q}(t) = Pr\{d_{Q,R}(t)=d_{Q,knn}\ |\ I_Q(t)\} 
\end{equation}

Second, we estimate the time $t_{Q}$ it takes to find the exact $k$-NN. As we already discussed, this time can be significantly faster than the time needed to complete the search. Let $\hat{t}_{Q}$ be its estimate. We express it as a probability density function:
\begin{equation}\label{eq:density-time}
\small
 r_{Q,t}(x) = Pr\{t_{Q} = x\ |\ I_Q(t)\} 
\end{equation}
which expresses the conditional probability that $t_{Q}$ is equal to $x$, given information $I_Q(t)$. From this, we derive its cumulative distribution function $R_{Q}(\cdot)$. 
Then, given a confidence level $1 - \phi$, we can find a probabilistic upper bound $\tau_{Q,\phi}$ such that $R_{Q}(\tau_{Q,\phi}) = 1 - \phi$; $\phi$ represents the probability that the progressive answer at time $\tau_{Q,\phi}$ is not the exact, i.e., the proportion of bounds that fail to include the exact answer.


\subsection{Stopping Criteria}
\label{sec:stopping}
Based on the provided estimates, users may decide to trust the current progressive result and possibly stop their search. Which \emph{stopping criterion} to use is not straightforward and depends on whether users prioritize guarantees about the $k$-NN distance, about the relative error of the current progressive result, or about the exact answer itself.


An analyst may choose to stop query execution as soon as the prediction interval of the $k$-NN distance lies above a low threshold value. 
Unfortunately, this strategy raises some concerns. 
Previous work on progressive visualization~\cite{micallef:2019} discusses the problem of confirmation bias, where an analyst may use incomplete results to confirm a ``preferred hypothesis''. 
This is a well-studied problem in sequential analysis~\cite{wald1945}. It relates to the multiple-comparisons problem~\cite{Zgraggen:2018} and is known to increase the probability of a Type I error (false positives). 
We evaluate how such multiple sequential tests affect the accuracy of our methods, but discourage their use as stopping criteria, and instead propose the following three.

A first approach is to make use of the relative distance error estimate $\hat{\epsilon}_Q(t)$ (see Eq.~\ref{eq:error}). For instance, the analyst may decide to stop the search when the upper bound of the error's interval is below a threshold $\epsilon = 1\%$. 
An error-based stopping criterion offers several benefits: 
(i) the choice of a threshold does not depend on the dataset, so its interpretation is easier; 
(ii) this criterion does not inflate Type I errors as long as the threshold $\epsilon$ is fixed in advance;  
(iii) the error $\epsilon_Q(t)$ monotonically converges to zero (the same holds for the bounds of its estimates), thus there is a unique point in time $\tau_{Q,\theta,\epsilon}$ at which the bound of the estimated error reaches $\epsilon$, where $1-\theta$ is our confidence level (here, $\theta$ determines the proportion of times for which the relative distance error of our result will be greater than $\epsilon$).

A second approach is to use the $\tau_{Q,\phi}$ bound (see Section~\ref{sec:exact}) to stop the search, which provides guarantees about the proportion of exact answers, rather than the distance error. 
It also depends on a single parameter, rather than two. 
To avoid the multiple-comparisons problem, we provide a single estimate of this bound at the very beginning of the search, allowing users to plan ahead their stopping strategy.

A third approach is to bound the probability $p_{Q}(t)$. Specifically, we stop the search when this probability exceeds a level $1-\phi$, where $\phi$ here represents the probability that the current progressive answer is not the exact. 
We experimentally assess the tradeoffs of these three stopping criteria.

\subsection{Family-wise Error in Progressive \emph{k}-NN Queries}
\label{sec:family-wise}
A $k$-NN similarity search query aims to identify $k$ data series as answers (not simply the $k$-th NN of $Q$). 
In most practical scenarios, an analyst is thus interested in stopping criteria that apply to all the progressive answers of a $k$-NN query. 

We observe that similarity search algorithms always find the exact $i$-NN to a query before its exact $(i+1)$-NN. Therefore, our second ($\tau_{Q,\phi}$) and third ($p_{Q}(t)$) stopping criteria naturally apply to all $k$ answers of a $k$-NN query. If the $k$-th answer is exact when the search stops, then we also know that answers of a rank lower are also exact.

In contrast, our first criterion on the relative distance error is optimistic. Stopping when the relative distance error $\epsilon_Q(t)$ of the $k$-th answer is lower than $\epsilon$ does not provide any guarantee about the relative distance error of lower-rank answers. To deal with this problem, we focus instead on the relative \emph{family-wise} distance error, defined as follows:
\begin{equation}\label{eq:family-error}
\small
 \epsilon^f_{Q}(t) = \frac{d_{Q,R}(t)}{d^f_{Q,knn}(t)} - 1
\end{equation}
where the distance term $d^f_{Q,knn}(t) \leq d_{Q,knn}$ represents a $k$-NN distance that is corrected for the family-wide error at time $t$, such that:
\begin{equation}\label{eq:correction}
d^f_{Q,knn}(t)  = \frac{d_{Q,knn}}{\max\limits_{1 \leq i \leq k} \{d_{Q,R_i}(t) / d_{Q,inn}\}}
\end{equation}
Our goal now is to find an estimate $\hat{d}^f_{Q,knn}(t)$.

\section{Prediction Methods}
\label{sec:methods}

We now present our approach, called \emph{ProS}.
We use 1 synthetic and 3 real datasets (i.e., seismic, SALD, and deep1B) from past studies~\cite{Zoumpatianos:2016,Echihabi:2018} to showcase our methods. 
We further explain and use these datasets in Section~\ref{sec:experiments} (see Table~\ref{tab:datasets} for a summary of their characteristics) to evaluate our methods.

Our goal is to support reliable prediction with small training sets of 
queries. We are also interested in expressing the uncertainty of our predictions with well-controlled bounds, as discussed in the previous section. We thus focus on statistical models that capture a small number of generic relationships in the data.
We first examine methods that assume constant information ($I_Q(t)=I_Q$). They are useful for providing an initial estimate \emph{before} the search starts. We distinguish between query-sensitive methods, which take into account the query series Q, and query-agnostic methods, which provide a common estimate irrespective of $Q$ ($I_Q=I$). Inspired by Ciacca et al.~\cite{Ciaccia:1998:CMS,Ciaccia:1999}, these methods serve as baselines to compare to a new set of progressive methods. Our progressive methods update information during the execution of a search, resulting in considerably improved predictions.



To simplify our analysis, we focus on $1$-NN similarity search. 
At the end of the section, we explain how our analysis naturally extends to $k$-NN search.

\subsection{Initial $1$-NN Distance Estimates}


We first concentrate on how to approximate the distribution function $h_{Q,0}(x)$ (see Equation~\ref{eq:density}), thus provide estimates before similarity search starts. 

As Ciaccia et al.~\cite{Ciaccia:1999}, we rely on witnesses, which are ``training'' query series that are randomly sampled from a dataset.
Unlike their approach, however, we do not use the distribution of raw pairwise distances $F_Q(\cdot)$. 
Instead, for each witness, we execute $1$-NN similarity queries with a fast state-of-the-art algorithm, such as iSAX2+~\cite{journal/kais/Camerra2014}, or DSTree \cite{Wang:2013:DDS}. This allows us to derive directly the distribution of $1$-NN distances and predict the $1$-NN distance of new queries. 


This approach has two main benefits. 
First, we use the tree structure of the above algorithms to prune the search space and reduce pre-calculation costs. 
Rather than calculating a large number of pairwise distances, we focus on the distribution of $1$-NN distances with fewer distance calculations. 
Second, we achieve reliable and high-quality approximation with a relatively small number of training queries ($\approx 100 - 200$) independently of the dataset size (we report and discuss these results in Section~\ref{sec:experiments}). 

\vspace{5pt} 
\noindent{\bf Query-Agnostic Model (Baseline).} 
Let $\mathcal{W} = \{W_j | j = 1.. n_w\}$ be a set of $n_w=|\mathcal{W}|$ witnesses randomly drawn from the dataset. 
We execute a $1$-NN similarity search for each witness and build their $1$-NN distance distribution. 
We then use this distribution to approximate the overall (query-independent) distribution of $1$-NN distances $g_n(\cdot)$ and its cumulative probability function $G_n(\cdot)$. 
This method is comparable to Ciaccia et al.~\cite{Ciaccia:1998:CMS} query-agnostic approximation method and serves as a baseline.

\vspace{5pt} 
\noindent{\bf Query-Sensitive Model.} 
Intuitively, the smaller the distance between the query and a witness, the better the $1$-NN of this witness predicts the $1$-NN of the query. 
We capture this relationship through a random variable that expresses the weighted sum of the $1$-NN distance of all $n_w$ witnesses:
\begin{equation}\label{eq:wit-distance}
\small
dw_Q = \sum\limits_{j=1}^{n_w}({a_{Q,j} \cdot d_{W_i,1nn}})
\end{equation}
\noindent
\fanis{Similar to Ciacca et al.~\cite{Ciaccia:1999}, we use weights $a_{Q,j}$ that decrease exponentially to the distance between the query $Q$ and the $j^{th}$ witness:} 
\begin{equation}\label{eq:weights}
\small
a_{Q,j} = \frac{d(Q, W_j)^{-exp}}{\sum\limits_{i=1}^{n_w}{d(Q, W_i)^{-exp}}}
\end{equation}
Our tests have shown optimal results for exponents $exp$ that are close to $5$. For simplicity, we use $exp = 5$ for all our analyses. 
Additional tests have shown that the fit of the model becomes consistently worse if witnesses are selected with the GNAT algorithm~\cite{Brin:1995,Ciaccia:1999} (we omit these results for brevity). Therefore, we only examine random witnesses here.

\begin{figure}[tb]
\centering
\footnotesize
\includegraphics[width=1.\linewidth]{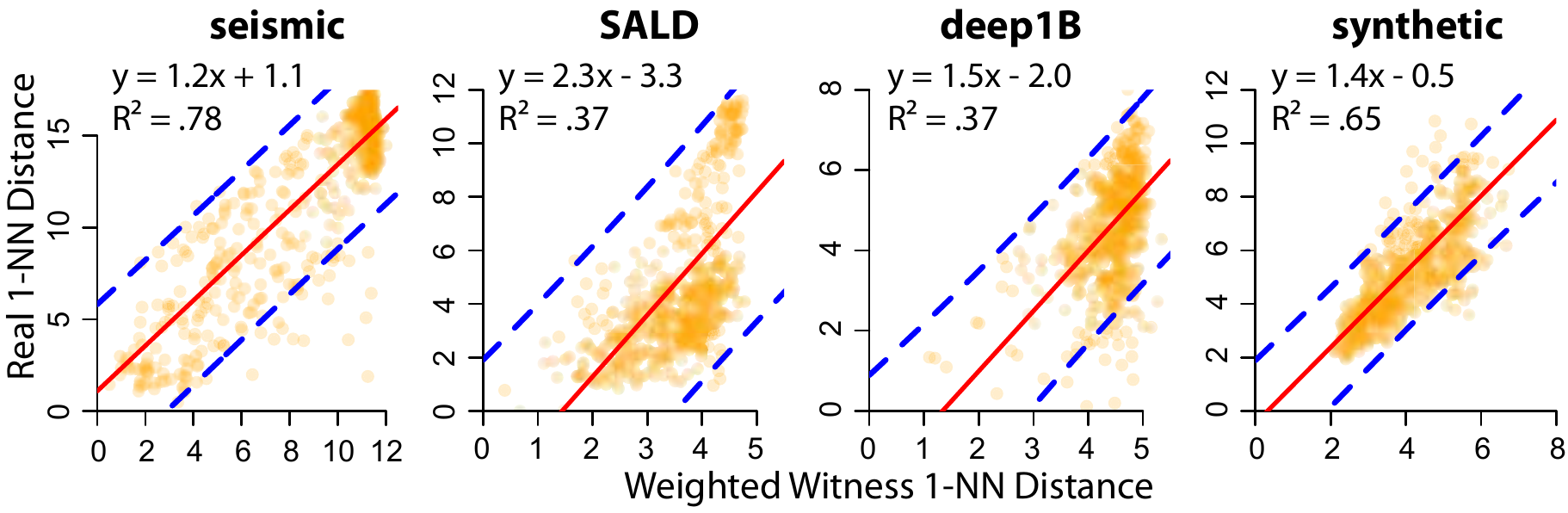}
\caption{Linear models (red lines) predicting the real 1-NN distance $d_{Q,1nn}$ based on the weighted witness 1-NN distance $dw_Q$ for $exp = 5$. All models are based on $n_w=200$ random witnesses and $n_r=100$ queries, and tested on 500 queries (in orange). The blue dashed lines show the range of their $95\%$ prediction intervals.}
\label{fig:wit-fit}
\end{figure}

We use $dw_Q$ as predictor of the query's real $1$-NN distance $d_{Q,1nn}$ and base our analysis on the following linear model: 
\begin{equation}\label{eq:wit-model}
 d_{Q,1nn}  = \beta \cdot dw_Q + c 
\end{equation}
Figure~\ref{fig:wit-fit} shows the parameters of instances of this model for the four datasets of Table~\ref{tab:datasets}. 
We conduct linear regressions by assuming that the distribution of residuals is normal (Gaussian) and has equal variance.


Since the model parameters ($\beta$ and $c$) and the variance are dataset specific, they have to be trained for each individual dataset. To train the model, we use an additional random sample of $n_r$ \emph{training queries} that is different from the sample of witnesses. Based on the distance of each training query $Q_i$ from the witnesses, we calculate $dw_{Q_i}$ (see Equation~\ref{eq:wit-distance}). We also run similarity search to find its $1$-NN distance $d_{Q_i,1nn}$. 
We then use all pairs $(dw_{Q_i}, d_{Q_i,1nn})$, where $i=1..n_r$, to build the model.
The approach allows us to construct both point estimates (see Equation~\ref{eq:wit-distance}) and prediction intervals (see Figure~\ref{fig:wit-fit}) that provide probabilistic guarantees about the range of the $1$-NN distance. 


\subsection{Progressive $1$-NN Distance Estimates}
\label{sec:distance-estimates}
So far, we have focused on initial $1$-NN distance estimates. 
Those do not consider any information about the partial results of a progressive similarity-search algorithm. 
Now, given Definition~\ref{def:progressive}, the distance of a progressive result $d_{Q,R}(t_i)$ will never deteriorate and thus can act as an upper bound for the real $1$-NN distance. 
The challenge is how to provide a probabilistic lower bound that is larger than zero.

\begin{figure}[tb]
\centering
\footnotesize
\includegraphics[width=0.99\linewidth]{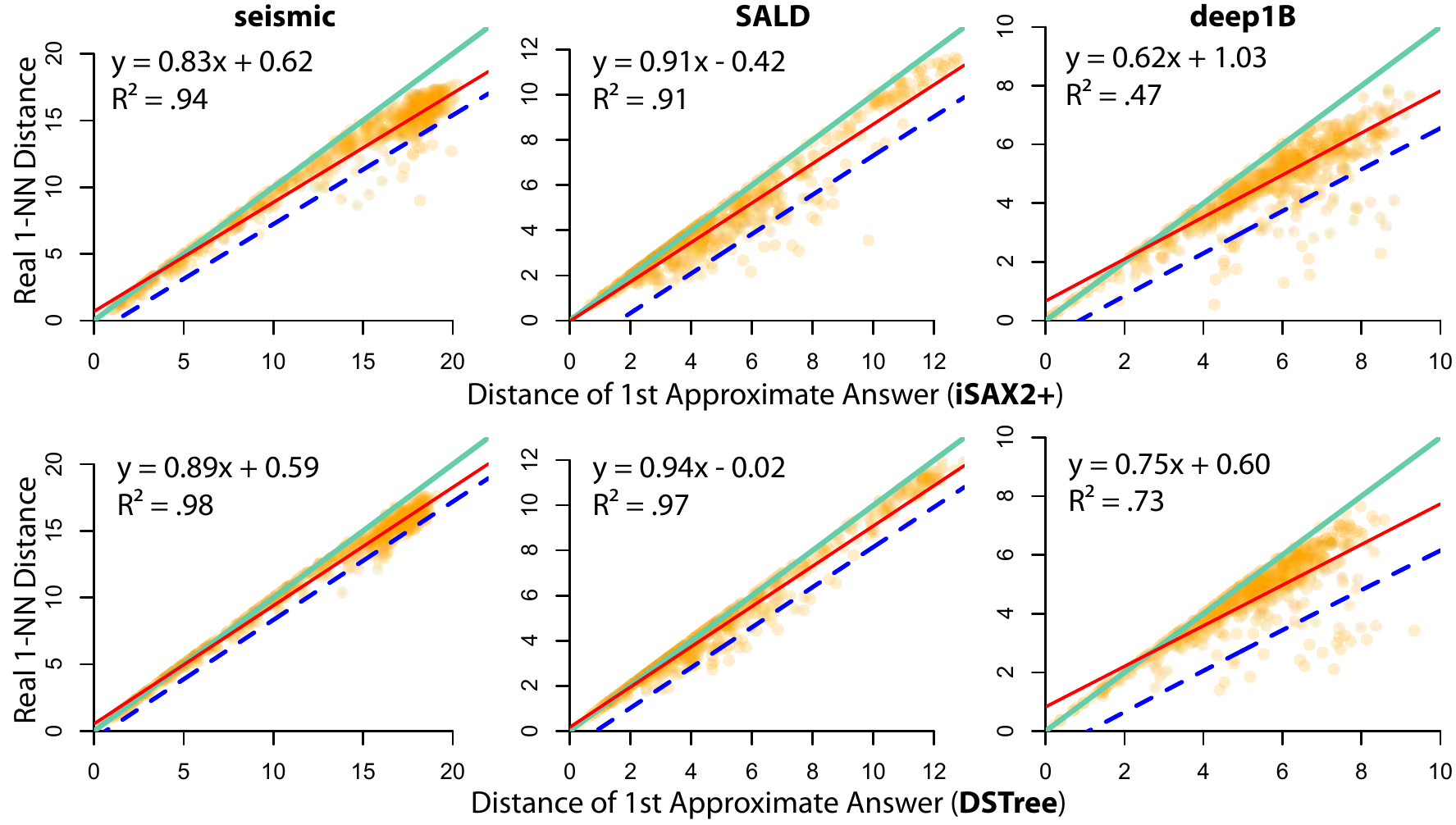}
\caption{Linear models (red solid lines) predicting the real $1$-NN distance $d_{Q,1nn}$ based on the \emph{first} approximate answer distance of iSAX2+ and DSTree. All models are trained with 200 queries. The 500 (orange) points in each plot are queries out of the training set. 
Green (solid) lines ($y = x$) are hard upper bounds, set by the approximate answer.
Blue lines show the range of one-sided $95\%$ prediction intervals that form probabilistic lower bounds.} 
\label{fig:bsf-fit}
\end{figure}

Our approach relies on the observation that the approximate answers of index-based algorithms are generally close to the exact answers. 
Figure~\ref{fig:bsf-fit} illustrates the relationship between the true $1$-NN distance and the distance of the first progressive (approximate) answer returned by iSAX2+~\cite{journal/kais/Camerra2014}. 
(The results for the DSTree index~\cite{Wang:2013:DDS} that follows a completely different design from iSAX2+ are very similar; we omit them for brevity).
We observe a strong linear relationship for both algorithms, especially for the DSTree index.
We can express it with a linear model and then derive probabilistic bounds in the form of prediction intervals. 
As shown in Figure~\ref{fig:bsf-fit}, the approach is particularly useful for constructing lower bounds. 
Those are clearly greater than zero and provide valuable information about the extent to which a progressive answer can be improved or not.

Since progressive answers improve over time and tend to converge to the $1$-NN distance, we could take such information into account to provide tighter estimates as similarity search progresses. 
To this end, we examine different progressive prediction methods. 
They are all based on the use of a dataset of $n_r$ training queries that includes information about all progressive answers of a similarity search algorithm to each query, including a timestamp and its distance.

\vspace{5pt} 
\noindent{\bf Linear Regression.} Let $t_1, t_2,...,t_m$ be specific moments of interest (e.g., 100ms, 1s, 3s, and 5s). Given $t_i$, we can build a time-specific linear model:
\begin{equation}\label{eq:individual-distance}
d_{Q,1nn}  = \beta_{t_i} \cdot d_{Q,R} (t_i)+ c_{t_i}
\end{equation}

\noindent
where $d_{Q,R}(t_i)$ is Q's distance from the progressive answer at time $t_i$. 
As an advantage, this method produces models that are well adapted to each time of interest.
On the downside, it requires the pre-specification of a discrete set of time points, which may not be desirable for certain application scenarios. 
However, building such models from an existing training dataset is inexpensive, so reconfiguring the moments of interest at use time is not a problem. 

The above model can be enhanced with an additional term $\beta \cdot dw_Q$ (see Equation~\ref{eq:wit-distance}) that takes witness information into account. 
However, this term results in no measurable improvements in practice, so we do not discuss it further.

\vspace{5pt} 
\noindent{\bf Kernel Density Estimation.} A main strength of the previous method is its simplicity.
However, linearity is a strong assumption that may not always hold. Other assumption violations, such as heteroscedasticity, can limit the accuracy of linear regression models. As alternatives, we investigate non-parametric methods that approximate the density distribution function $h_{Q,t}(\cdot)$ based on multivariate kernel density estimation~\cite{wand:1993,tarn:2005}. 

As for linear models, we rely on the functional relationship between progressive and final answers. We represent this relationship as a 3-dimensional density probability function $k_Q(x, y, t)$ that expresses the probability that the $1$-NN distance from $Q$ is $x$, given that $Q$'s distance from the progressive answer at time $t$ is $y$. From this function, we derive $h_{Q,t}(x)$ by setting $y = d_{Q,R}(t)$.

We examine two approaches for constructing the function $k_Q(\cdot, \cdot, \cdot)$. As for linear models, we specify discrete moments of interest $t_i$ and then use bivariate kernel density estimation~\cite{wand:1994} to construct an individual density probability function $k_Q(\cdot, \cdot, t_i)$. Alternatively, we construct a common density probability function by using 3-variate kernel density estimation. The advantage of this method is that it can predict the $1$-NN distance at any point in time. 
Nevertheless, this comes with a cost in terms of precision (see Section~\ref{sec:experiments}). 

The accuracy of kernel density estimation highly depends on the method that one uses to smooth out the contribution of points (2D or 3D) in a training sample. We use gaussian kernels, but for each estimation approach, we select bandwidths with a different technique. We found that the plug-in selector of Wand and Jones~\cite{wand:1994} works best for our bivariate approach, while the smoothed cross-validation technique~\cite{tarn:2005} works best for our 3-variate approach.

\vspace{5pt} 
\noindent{\bf Measuring Time.} So far, we have based our analysis on time. 
Nevertheless, time (expressed in seconds) is not a reliable measure for training and applying models in practice. 
The reason is that time largely depends on the available computation power, which can vary greatly across different hardware settings. 
Our solution is to use alternative measures that capture the progress of computation without being affected by hardware and computation loads. One can use either the number of series comparisons (i.e., the number of distance calculations), or the number of visited leaves. Both measures can be easily extracted from the iSAX2+~\cite{journal/kais/Camerra2014}, the DSTree~\cite{Wang:2013:DDS}, or other tree-based similarity-search algorithms. 
Our analyses in this paper are based on the number of visited leaves (\emph{Leaves Visited}). We should note that for a given number of visited leaves, we only consider a single approximate answer, which is the best-so-far answer after traversing the last leaf.

\subsection{Estimates for Exact Answers}
We investigate two types of estimates for exact answers (see Section~\ref{sec:exact}): (i) progressive estimates of the probability $p_{Q}(t)$ that the $1$-NN has been found; and (ii) query-sensitive estimates of the time $t_{Q}$ that it takes to find the exact answer. 
We base our estimations on the observation that queries with larger $1$-NN distances tend to be harder, i.e., it takes longer to find their $1$-NN. Now, since approximate distances are good predictors of their exact answers (see previous subsection), we can also use them as predictors of $p_{Q}(t)$ and $t_{Q}$.

\begin{figure}[tb]
\centering
\footnotesize
\includegraphics[width=1\linewidth]{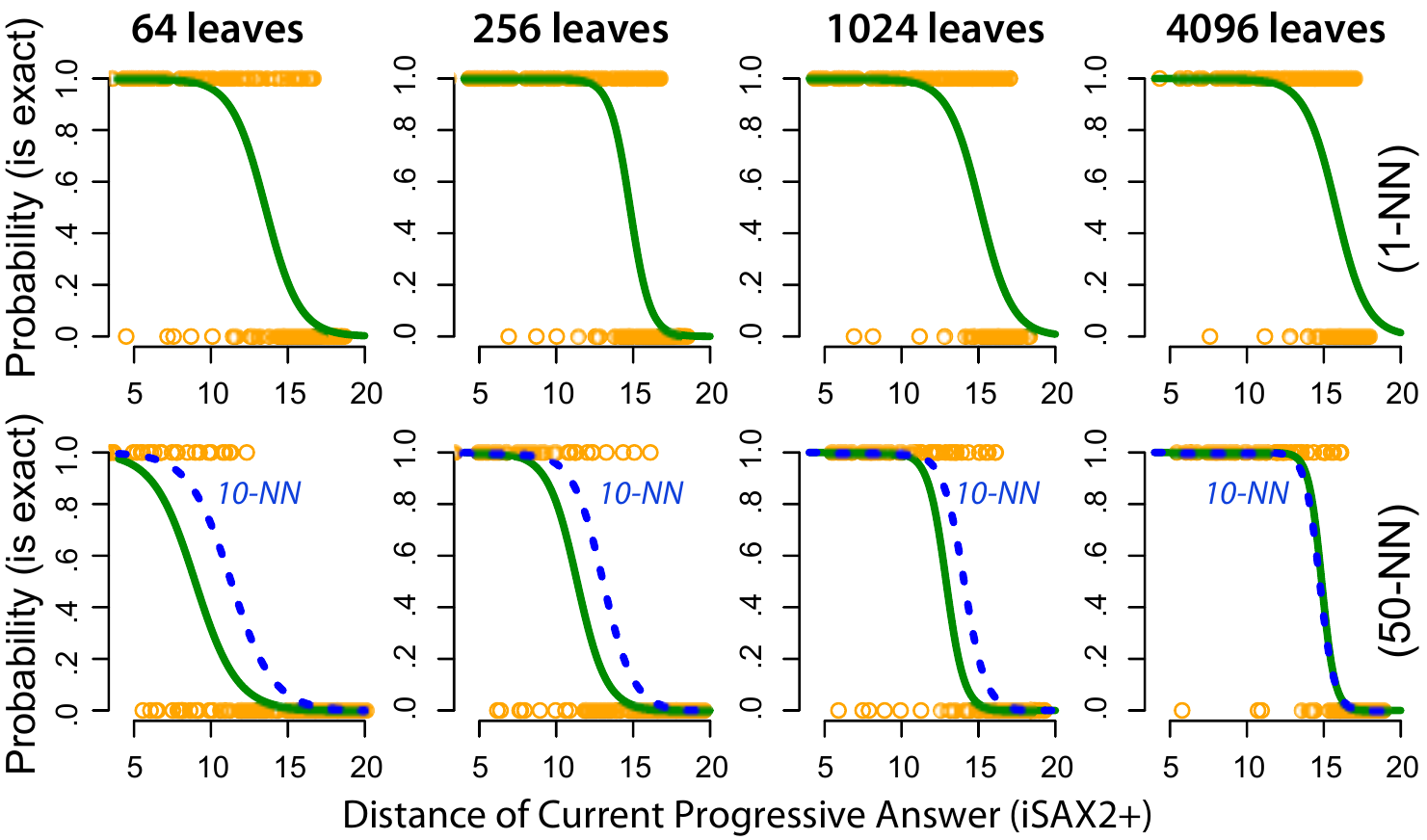}
\caption{Estimating the probability of exact answers with 100 training queries based on the current $1$-NN (top) and $50$-NN (bottom) progressive answer (seismic dataset). We show the logistic curves (in green) at different points in time (64, 256, 1024, and 4096 leaves) and 200 test queries (in orange). For comparison, we also show the logistic curve for the 10-NN (blue dashed line).}
\label{fig:logistic}
\end{figure}

\vspace{5pt} 
\noindent{\bf Probability Estimation.} Let $t_1, t_2,...,t_m$ be moments of interest, and let $d_{Q,R}(t_i)$ be the distance of the progressive result at time $t_i$. We use logistic regression to model the probability $p_{Q}(t_i)$ as follows: 
\begin{equation}\label{eq:probability}
\small
ln{\frac{p_{Q}(t_i)}{1 - p_{Q}(t_i)}} = \beta_{t_i} \cdot d_{Q,R}(t_i) + c_{t_i}
\end{equation}
Again, we can use the number of visited leaves to represent time. 
Figure~\ref{fig:logistic} presents an example for the seismic dataset, where we model the probability of exact answers for four points in time (when 64, 256, 1024, and 4096 leaves are visited). 
We observe that over time, the curve moves to the right range of distances, and thus, probabilities increase.   

Note that we have considered other predictors as well (such as the time passed since the last progressive answer), but they did not offer any predictive value.

\begin{figure}[tb]
\centering
\footnotesize
\includegraphics[width=1\linewidth]{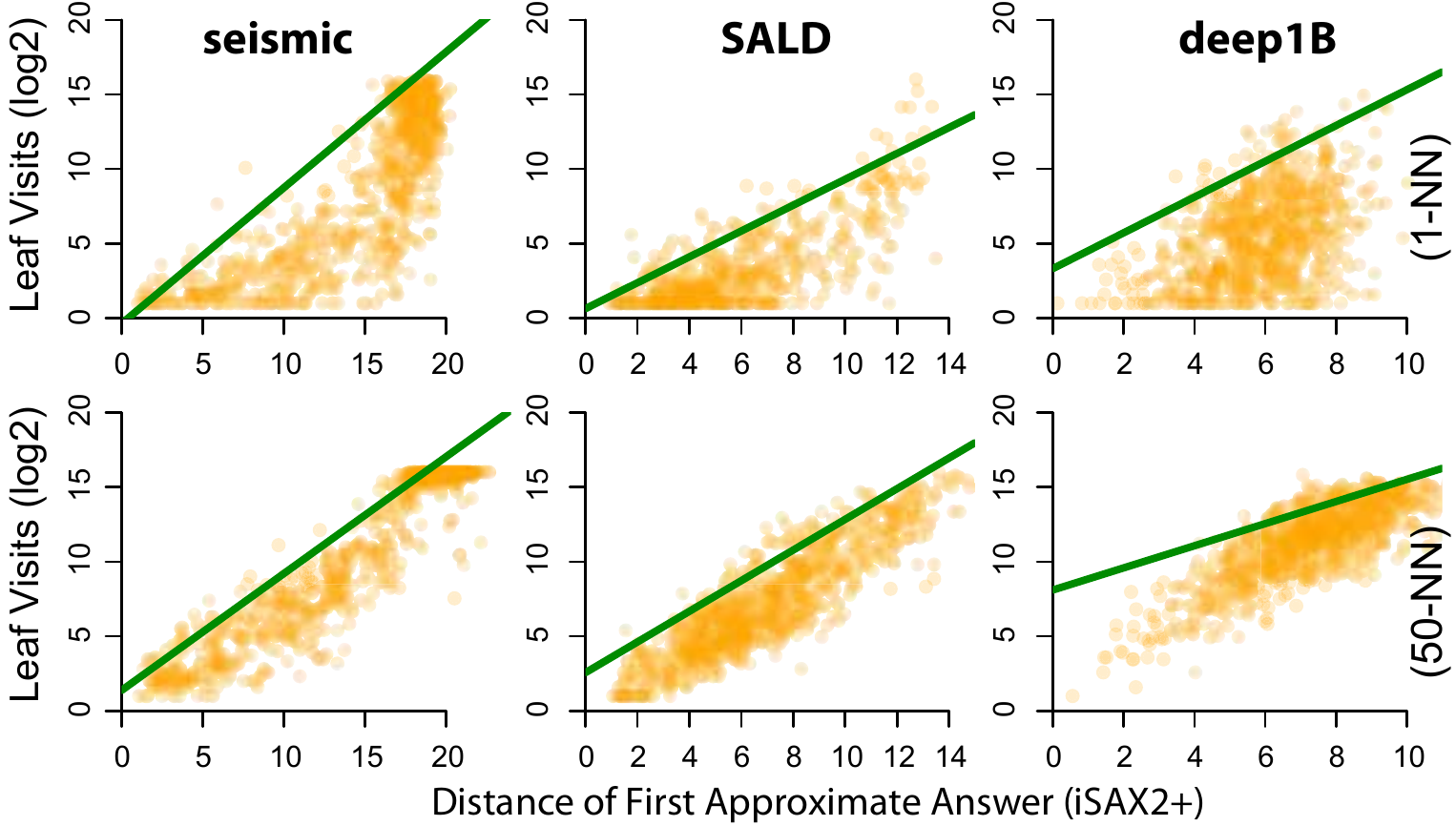}
\caption{Upper time bounds ($\phi = .05$) for 1-NN (top) and 50-NN (bottom) answers. Bounds (in green) are constructed from 100 random queries through quantile regression, where we estimate the $95\%$ quantile of the logarithm of leaf visits as a function of the distance of the 1st approximate answer.}
\label{fig:quantile}
\vspace{-12pt}
\end{figure}

\vspace{5pt} 
\noindent{\bf Time Bound Estimation.} As we explained in Section~\ref{sec:stopping}, we provide a single estimate for $t_{Q}$ at the very beginning of the search. Figure~\ref{fig:quantile} (top) illustrates the relationship between the distance of the first approximate answer and the number of leaves (in logarithmic scale) at which the $1$-NN is found. We observe that the correlation between the two variables is rather weak. However, we can still extract meaningful query-sensitive upper bounds, shown as green lines.  
To construct such bounds, we use quantile regression~\cite{koenker2005}. This method allows us to directly estimate the $1 - \phi$ quantile of  the time needed to find the exact answer, i.e., derive the upper bound $\tau_{Q,\phi}$. As a shortcoming, the accuracy of quantile regression is sensitive in small samples. Nevertheless, we show that 100 training queries are generally enough to ensure high-quality results.

\fanis{
\vspace{5pt} 
\noindent{\bf Example.}} Figure~\ref{fig:examples-vis} presents an example that illustrates how the above methods can help users assess how far from the $1$-NN their current answers are. We use a variation of pirate plots~\cite{yarrr} to visualize the $1$-NN distance estimate $\hat{d}_{Q,1nn}(t)$ and the relative error estimate $\hat{\epsilon}_Q(t)$ by showing their probability density distribution and their $95\%$ prediction interval. We also communicate the probability $p_{Q}(t_i)$ and a probabilistic bound $\tau_{Q,\phi}$ ($\phi = .05$) after the first visited leaf. 
\fanis{The initial distance estimate based on witnesses is rather wide. However, prediction intervals become tighter as soon as search starts.    
In particular, }the upper bound of the error estimate drops below $10\%$ within 1.1sec, while the probability that the current answer is exact is estimated as $98\%$ after 15.7sec (total query execution time for this query is 75.2sec). In this example, the $1$-NN is found in 3.8sec. 
%

\subsection{Progressive Estimates for $k$-NN Similarity Search}
\label{sec:knn-search}

The predictions methods presented above naturally extend to the general case of $k$-NN search. As Figure~\ref{fig:logistic} shows, exact answers for larger $k$ ranks are found later in time. Still, distance is a good predictor of whether a progressive answer is exact. We observe that at earlier steps, uncertainty is higher for large $k$ ranks, but as more leaves are visited, the prediction quality of the logistic model improves.

Figure 5 (bottom) presents how upper time-bound estimation extends to $50$-NN. We can still derive useful bounds based on the distance of the very first approximate answer. Interestingly, the correlation between this distance and the logarithm of visited leaves is stronger now. We could eventually use this behavior to construct meaningful lower time bounds for $k$-NN search.

For $k$-NN search, we evaluate the family-wise error of distance estimates $\hat{\epsilon}^f_{Q}(t)$ based on Equation~\ref{eq:family-error}. We apply the same prediction methods (see Section~\ref{sec:distance-estimates}) but our dependent variable is now $d^f_{Q,knn}$ (rather than $d_{Q,1nn}$). We use Equation~\ref{eq:correction} to calculate $d^f_{Q,knn}$ for our training dataset.

\subsection{Dynamic Time Warping (DTW)}
\label{section:dtw}

The data series indexes we employ, i.e., iSAX2+ and DSTree, originally supported only the Euclidean distance. 
We modified their query answering algorithms to provide support for DTW based on the ideas proposed in~\cite{keogh2005exact}. 

First, we find $U$ and $L$, the upper and lower envelopes that bound the query $Q$ according to the Sakoe-Shiba constraint~\cite{sakoe-shiba} using the algorithm proposed in~\cite{DBLP:journals/pr/Lemire09}. 
Then, for each index, we calculate $\hat{U}$ and $\hat{L}$ the summarizations of $U$ and $L$, and we derive $MinDist(Q,N)$, the lower bounding distance between the summarized envelopes $\hat{U}$ and $\hat{L}$ of the query $Q$ and an index node $N$. 
Note that we probe the index using the summarized envelopes $\hat{U}$ and $\hat{L}$ rather than the query $Q$ itself. The distance $MinDist(Q,N)$ is guaranteed to lower bound $LB_{Keogh}$\footnote{We note that other lower bounds for DTW can be used as well, such as LB\_Improved~\cite{DBLP:journals/pr/Lemire09}. Even though LB\_Improved can produce tighter bounds, previous experiments have resulted in higher query answering times due to the additional computations it involves~\cite{messijournal}.}, which itself lower bounds $DTW$. 
\begin{equation}\label{eq:lb-keogh}
\small
LB_{Keogh}(Q,C) = \sqrt{\sum_{i=1}^{n}\begin{cases}
	({c_i}-{U_i})^2&\text{if  }{c_i}>{U_i}\\    
	({L_i}-{c_i})^2&\text{if  }{c_i}<{L_i}\\
	0 &otherwise
	\end{cases}}
\end{equation}

\begin{figure*}[!htb]
	\captionsetup[subfigure]{justification=centering}
	\centering
	\begin{subfigure}{0.32\textwidth}
		\centering
		\includegraphics[width=\textwidth]{{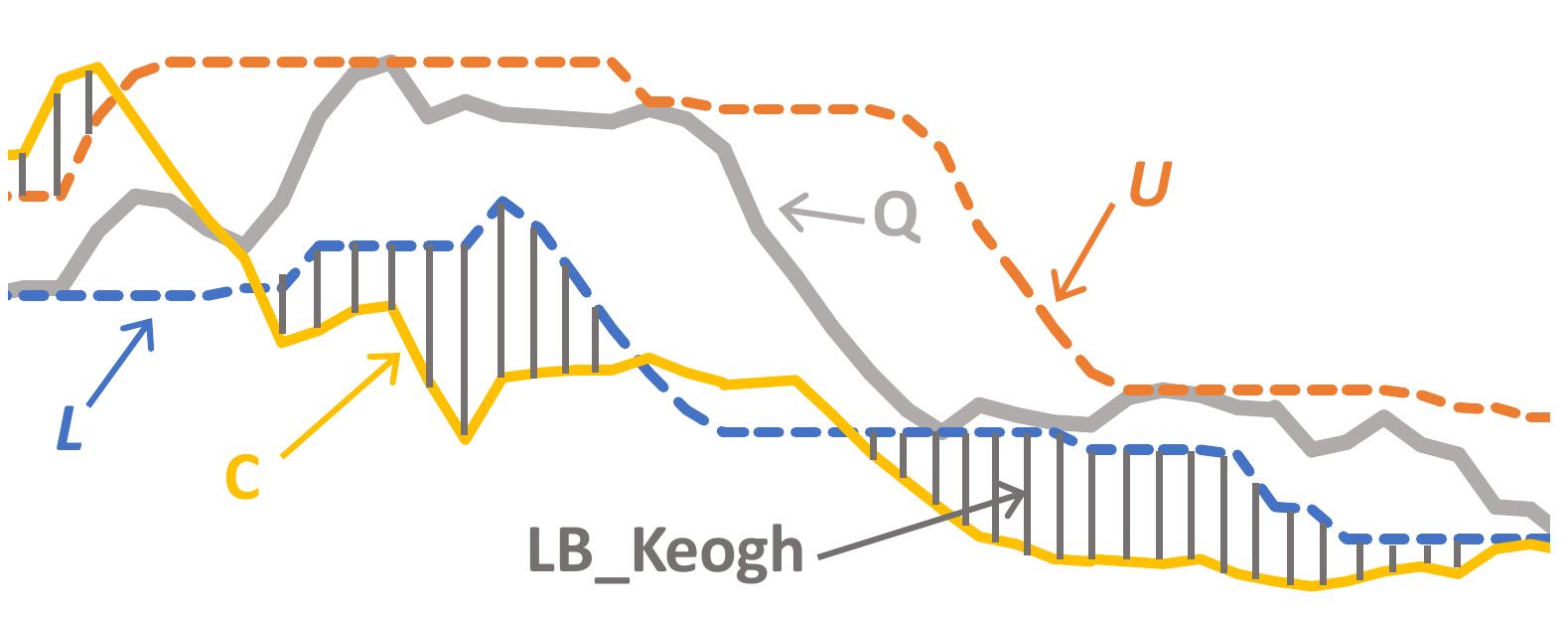}}
		\caption{$LB_{Keogh}$}
		\label{fig:envelope-keogh}
	\end{subfigure}
	\begin{subfigure}{0.32\textwidth}
		\centering
		\includegraphics[width=\textwidth]{{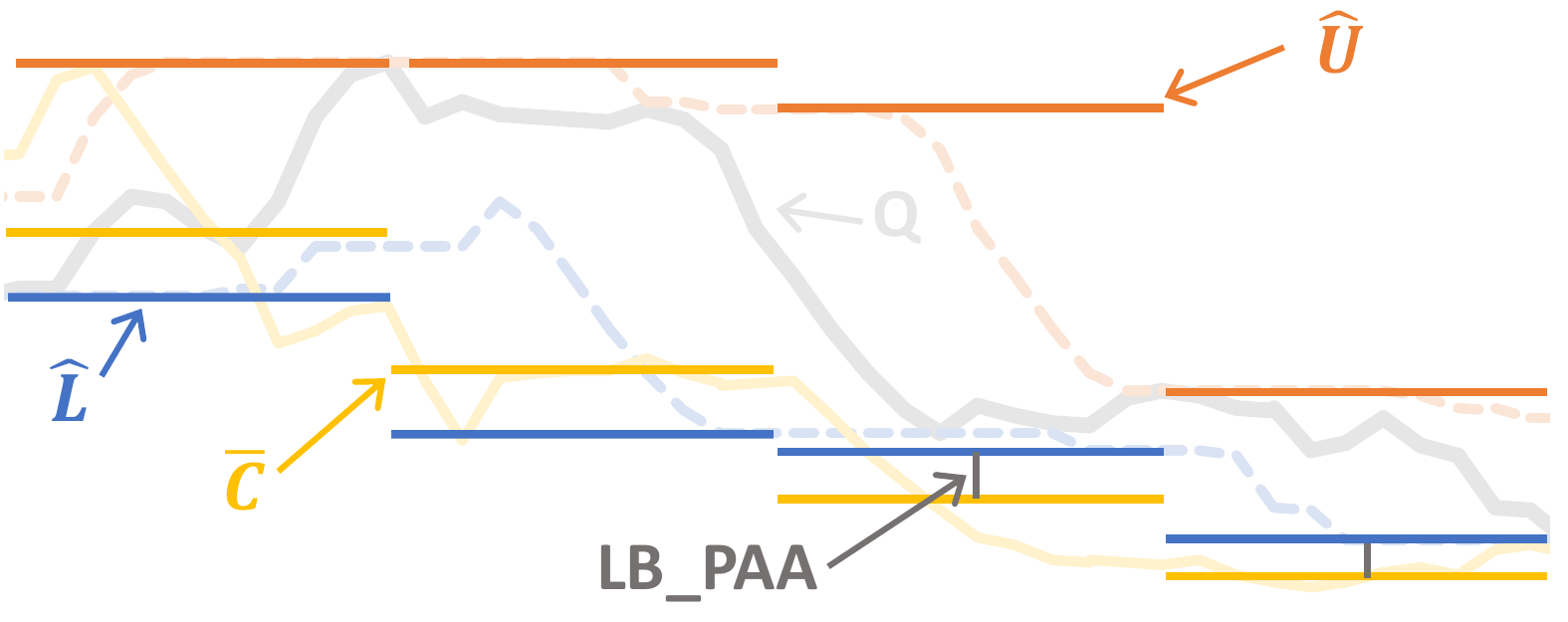}}
		\caption{$LB_{PAA}$}
		\label{fig:envelope-paa}
	\end{subfigure}
	\begin{subfigure}{0.32\textwidth}
		\centering
		\includegraphics[width=\textwidth]{{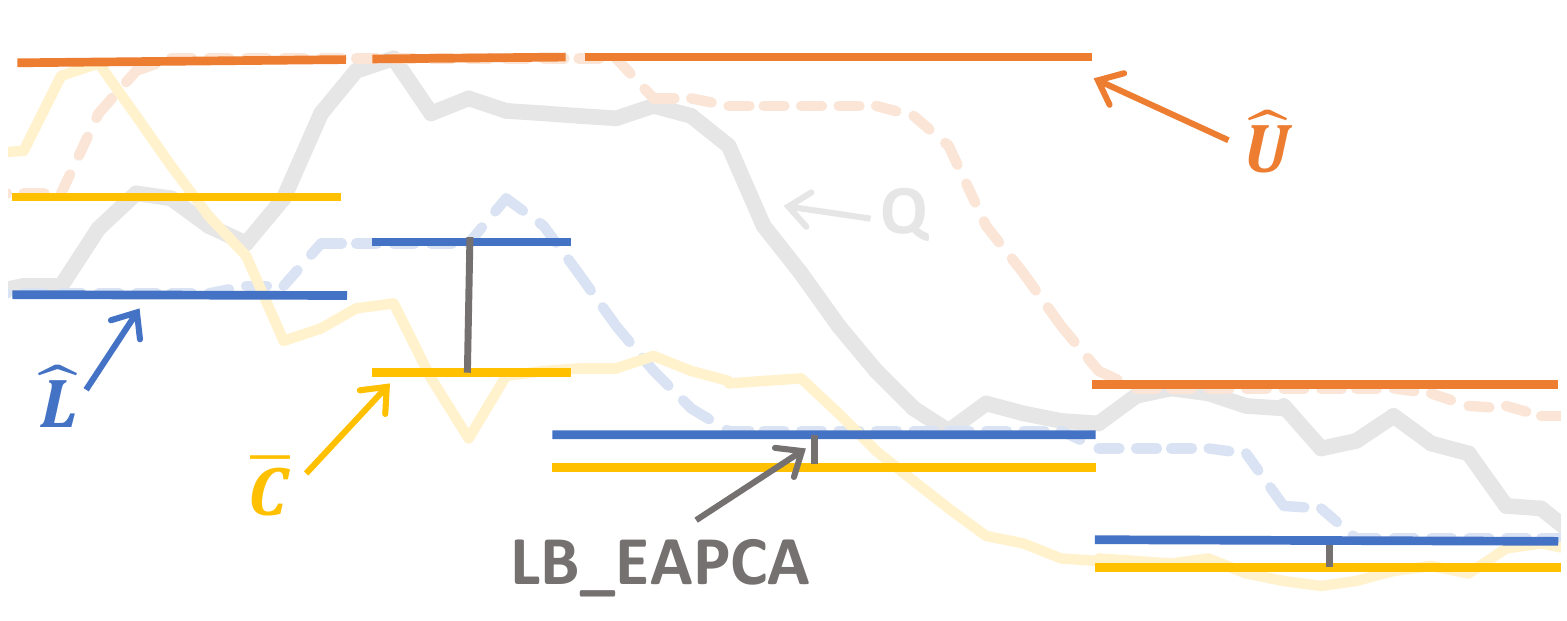}}
		\caption{$LB_{EAPCA}$}
		\label{fig:envelope-eapca}
	\end{subfigure}
	\caption{Envelopes $U$ and $L$ of query Q with warping size 10\%. Each dark gray vertical line contributes to the lower bounding distance between $Q$ and a candidate answer $C$. Distances are calculated by taking the square root of the sum of the squares of the lengths of the gray lines. In the case of $LB_{PAA}$ and $LB_{EAPCA}$, the squares of the lengths are scaled by the number of points in each segment.}
	\label{fig:envelope}
\end{figure*}

We do not calculate $LB_{Keogh}$ directly because it is also computationally expensive. As Equation~\ref{eq:lb-keogh} and Figure~\ref{fig:envelope-keogh} indicate, it requires calculating distances between the individual $n$ high-dimensional points of the candidate $C$ and the envelopes $U$ and $L$ (if some points of the candidate fall inside the query envelope, then their distance is zero). On the other hand, 
$MinDist(Q,N)$ is faster to compute because it consists of the distances between the segments of a node $N$ and the segments of the summarized envelopes. We use $\bar{C}$, $\bar{Q}$, $\hat{U}$ and $\hat{L}$ to refer to the summaries of $C$, $Q$, $L$ and $U$ respectively. The specific representation used for the summaries can be inferred from the context.

\noindent{\bf iSAX2+.}
Since iSAX2+ is based on $PAA$, we use the same formulas as those in~\cite{keogh2005exact} to derive $\hat{U}$ and $\hat{L}$ (Equations~\ref{eq:u-hat-paa}-\ref{eq:l-hat-paa}), the special piecewise aggregate approximations of $U$ and $L$ for the $i^{th}$ segment of $\bar{Q}$, and calculate $MinDist_{PAA}(Q,N)$ (Equation~\ref{eq:lb-paa}). We consider a PAA summarization using $M$ segments.
\begin{equation}\label{eq:u-hat-paa}
\small
\hat{U}_i = max(U_{(\frac{n}{M})(i-1)+1},\ \dots\ ,\ U_{(\frac{n}{M})(i)})
\end{equation}
\begin{equation}\label{eq:l-hat-paa}
\small
\hat{L}_i = max(L_{(\frac{n}{M})(i-1)+1},\ \ \dots\ ,\ L_{(\frac{n}{M})(i)})
\end{equation}

\begin{equation}\label{eq:lb-paa}
\scriptsize
LB_{PAA}(Q,\bar{C}) = \sqrt{\frac{n}{M}}\sqrt{\sum_{i=1}^{M}\begin{cases}
	(\bar{c}_i-\hat{U}_i)^2&\text{if  }\bar{c}_i>\hat{U}_i\\    
	(\hat{L}_i-\bar{c}_i)^2&\text{if  }\bar{c}_i<\hat{L}_i\\
	0 &otherwise
	\end{cases}}
\end{equation}

\begin{equation}\label{eq:MinDist-paa}
\scriptsize
MinDist_{PAA}(Q,N) = \sqrt{\frac{n}{M}}\sqrt{\sum_{i=1}^{M}\begin{cases}
	({l_i}-\hat{U}_i)^2&\text{if  }{l_i}>\hat{U}_i\\    
	(\hat{L}_i-{h_i})^2&\text{if  }{h_i}<\hat{L}_i\\
	0 &otherwise
	\end{cases}}
\end{equation}
Such that $l_i$ and $h_i$ are the lower and higher endpoints of the major diagonal of $R$, the smallest rectangle that spatially contains the $i$th segment of all data series in $N$. Figure~\ref{fig:envelope-paa} illustrates the $LB_{PAA}$ distance between $Q$ and $\bar{C}$.

\vspace{5pt} 
\noindent{\bf DSTree}. For DSTree, we propose new upper and lower bounding envelopes, $\hat{U}$ and $\hat{L}$, and a new lower bounding distance $MinDist_{EAPCA}(Q,N)$. Given an EAPCA representation with $M$ segments and $m_i$ is the right endpoint of segment $i$ $(m_1 <...< m_i < ...< m_M = n)$, 
the upper and lower EAPCA envelopes of segment $i$ of $\bar{Q}$ are defined as follows: 
\begin{equation}\label{eq:u-hat-eapca}
\small
\hat{U}_{i} = max(U_{m_{i-1}+1},\ U_{m_{i-1}+2},\ \dots\ ,\ U_{m_{i}})
\end{equation}
\begin{equation}\label{eq:l-hat-eapca}
\small
\hat{L}_{i} = min(L_{m_{i-1}+1},\ L_{m_{i-1}+2},\ \dots\ ,\ L_{m_{i}})
\end{equation}

The EAPCA lower bounding distance between $Q$ and $\bar{C}$ is defined as:


\begin{equation}\label{eq:lb-eapca}
\small
LB_{EAPCA}(Q,\bar{C}) = \sqrt{\sum_{i=1}^{M}(m_i-m_{i-1}) \ a_{i}}
	\end{equation}
	
	Where:
	\begin{equation}\label{eq:lb-eapca-a}
	\small
	a_{i} = \begin{cases}
	(\bar{c}_i-\hat{U}_i)^2&\text{if  }\bar{c}_i>\hat{U}_i\\    
	(\hat{L}_i-\bar{c}_i)^2&\text{if  }\bar{c}_i<\hat{L}_i\\
	0 &otherwise
	\end{cases}
\end{equation}

Figure~\ref{fig:envelope-eapca} shows the $LB_{EAPCA}$ distance between $Q$ and $\bar{C}$. The proof that $LB_{EAPCA}(Q,\Bar{C}) \leq LB\_Keogh(Q,C)$ is a straightforward extension of the proof in Proposition 1 in~\cite{keogh2005exact}.

Consider an EAPCA index node $N$ containing a set of data series $Y_1,\dots,Y_l$ with synopsis $\textbf{Z} = (z_1,z_2,\dots,z_M)$ where $z_i = (\mu_i^{min},\mu_i^{max})$ where $\mu_i^{min} = min(\mu_i^{Y_1},\dots,\mu_i^{Y_l})$ and $\mu_i^{max} = max(\mu_i^{Y_1},\dots,\mu_i^{Y_l})$. Then, the lower bounding distance between $Q$ and node $N$ is defined as:
\begin{equation}\label{eq:MinDist-eapca}
\small
MinDist_{EAPCA}(Q,N) = \sqrt{\sum_{i=1}^{M}(r_i-r_{i-1})(LB_i)}
	\end{equation}
	Such that	
	\begin{equation}\label{eq:MinDist-pca}
\small
LB_i = \begin{cases}
(\mu_i^{min}-\hat{U}_i)^2&\text{if  }\mu_i^{min}>\hat{U}_i\\    
(\hat{L}_i-\mu_i^{max})^2&\text{if  }\mu_i^{max}<\hat{L}_i\\
0 &otherwise
\end{cases}
\end{equation}

The proof that $MinDist_{EAPCA}(Q,N) \leq$\\ $LB_{EAPCA}(Q,\Bar{Y_j})$ $\forall\ 1\leq j \leq l$ is a straightforward extension of the proof in Theorem 2 in~\cite{Wang:2013:DDS}.
	
Note that although the DSTree exploits the standard deviation of the points in each segment to produce a tighter lower bound between a query $Q$ and a node $N$, we only use the mean. 
The reason is that the standard deviations of the points belonging to each segment of the EAPCA upper and lower envelopes are zero, and thus cannot contribute to the lower bound.

\section{Progressive $k$-NN Classification}
\label{sec:knn-classification-approach}

For $k$-NN classification, we can use again a progressive similarity search algorithm. At any given time, we take its progressive answer and use it to infer the \emph{progressive class.} 
However, since we are now interested in the class of the data series that serves as query, the notion of ``approximation'' is not relevant anymore -- the class can be either correct or wrong. 
A progressive answer in this case is only interesting if it returns the correct class, or alternatively, if it returns the same class (correct or not) as the non-progressive algorithm, i.e., the final, answer. 
In this case, we say that the class is the \emph{exact class}.

Our goal is now to provide guarantees for this class, rather than for the distance of the progressive answer. 
More formally, given a data series $Q$ as query, we run again a progressive $k$-NN search. At each time $t$, we take the most common class $c_Q(t)$ among the progressive $k$ nearest neighbors of $Q$. We then assess the probability $p_{c_Q}(t)$ that the class $c_Q$ of the exact answer is found, where as for $k$-NN similarity search, we use information $I_Q(t)$:
\begin{equation}\label{eq:probability_class}
\small
 p_{c_Q}(t) = Pr\{c_Q(t) \equiv c_Q\ |\ I_Q(t)\} 
\end{equation}

\noindent
Extending our ProS approach, we describe two solutions on how to either bound, or directly estimate this quantity.

\subsection{Bounding the Probability of Exact Class}
We can easily infer that $p_{c_Q}(t) \ge p_Q(t)$ (see Equation~\ref{eq:probability}), i.e., the probability that the current progressive class is exact is at least as high as the probability that the current progressive $k$-NN is exact. In other words, although the similarity search algorithm may have not yet found the exact answer to the $k$-NN similarity search query, the class can be the exact. 

A direct implication of the above is that the exact-answer probabilistic guarantees that we presented in Section~\ref{sec:exact} can be also considered as guarantees for the exact class. Likewise, the probability and time-bound stopping criteria presented in Section~\ref{sec:stopping} can also apply as stopping criteria for $k$-NN classification. Nevertheless, they are stricter, more conservative and result in reduced time savings. Instead, we update the stopping criteria by simply replacing the parameter $\phi$ by $\phi_c$, where $\phi_c$ represents the probability that the current progressive class is not the exact.

\subsection{Estimating the Probability of Exact Class}


We consider again $m$ moments of interest $t_1, t_2,..., t_m$. At each moment $t_i$, we estimate $p_{c_Q}(t_i)$ by using three predictors: (i) the distance $d_{Q,R}(t_i)$ of the $k$-NN, (ii) the current class $c_Q(t_i)$, and (iii) the extent to which the current $k$ answers agree on this class. The latter is quantified as follows: 
\begin{equation}\label{eq:prob-class}
\small
a(t_i) = \frac{n_{c_Q(t_i)} - 1}{k - 1}
\end{equation}
where $n_{c_Q(t_i)}$ is the number of occurrences of $c_Q(t_i)$ among the $k$ nearest neighbors returned by the progressive search ($k > 1$). We can the use these predictors to build a linear logistic regression model as in Equation~\ref{eq:probability}. 

Note that not all three predictors are always relevant. For example, if the number of available classes is large, information about the current class has no predictive value unless we use a much larger set of training queries. 
We have tested additional variables, such as the ones that evaluate the stability of $c_Q(t_i)$ over time, but we did not find them to be good predictors.


\section{Experimental Evaluation}
\label{sec:experiments}




\noindent{\bf Environment.}
All experiments were run on a Dell T630 rack server with two Intel Xeon E5-2643 v4 3.4Ghz CPUs, 512GB of RAM, and 3.6TB (2 x 1.8TB) HDD in RAID0. 

\vspace{5pt} 
\noindent{\bf Implementation.}  Our estimation methods were implemented in R. We use R's \emph{lm} function to carry our linear regression, the \emph{ks} library~\cite{ks} for multivariate kernel density estimation, and the \emph{quantreg} library~\cite{quantreg} for quantile regression. 
We use a grid of $200 \times 200$ points to approximate a 2D density distribution and a grid of $60 \times 180 \times 180$ points to approximate a 3D density distribution. 
Source code and datasets are in~\cite{webpage}.

\vspace{5pt} 
\noindent{\bf Datasets.}
For the evaluation of the progressive similarity search techniques, we used 1 synthetic and 3 real datasets from past studies~\cite{Zoumpatianos:2016,Echihabi:2018}, as well as an additional real dataset, PhysioNet~\cite{PhysioNet}. 
All datasets are \emph{100GB} in size with cardinalities and lengths reported in Table~\ref{tab:datasets}. 
\fanis{For our experiments with DTW, however, we used a smaller subset of these datasets (\emph{10GB} in size), since running them on the original datasets was extremely expensive.   }

\begin{table}
	\centering
	\small
	\caption{Experimental datasets for similarity search.}
	\scalebox{0.88}{
		\begin{tabular}{ l|c|c|c} 
			Name&Description&Num of series&Series length\\ \hline
			1. synthetic & random walks & 100M/10M & 256\\
			2. seismic~\cite{url/data/seismic} & seismic records & 100M/10M & 256\\ 
			3. SALD~\cite{url/data/sald} & MRI data & 200M/20M & 128\\ 
			4. deep1B~\cite{url/data/deep1b} & image descriptors & 267M/27M & 96\\ 
			\fanis{5. PhysioNet~\cite{PhysioNet}} & ECG recordings & \karima{20M/10M} & 256
		\end{tabular}
		\label{tab:datasets}
	}
	\vspace*{-0.3cm}
\end{table}

Synthetic data series were generated as random walks (cumulative sums) of steps that follow a Gaussian distribution (0,1). 
This type of data has been extensively used in the past~\cite{conf/sigmod/Faloutsos1994,journal/kais/Camerra2014,conf/kdd/Zoumpatianos2015} and models the distribution of stock market prices~\cite{conf/sigmod/Faloutsos1994}. 
The IRIS seismic dataset~\cite{url/data/seismic} is a collection of seismic instrument recordings from several stations worldwide (100M series of length 256). 
The SALD neuroscience dataset~\cite{url/data/sald} contains MRI data (200M series of length 128). 
The image processing dataset, deep1B~\cite{url/data/deep1b}, contains vectors extracted from the last layers of a convolutional neural network (267M series of length 96). \karima{The PhysioNet dataset~\cite{PhysioNet} contains ECG data (20 million series of length 256).} 


The above datasets are not annotated. In order to evaluate the progressive $k$-NN classification techniques, we use the datasets in Table~\ref{tab:datasets-classification}.

The Cylinder-Bell-Funnel (CBF) dataset~\cite{Saito:2000} is a synthetic dataset that has been used extensively in the data series classification community, consisting of data series belonging to one of three classes: cylinders, bells and funnels. 
Instances of each class are generated randomly with Gaussian noise added, such that each series has a fixed length of 128, but the onset and duration of each pattern varies randomly. 
An amplitude parameter is also used to control the difficulty of the dataset, where the smaller the amplitude, the less distinct the data series in different classes, thus the harder the classification task. 
We used the amplitude values 1 and 3 to generate the CBF1 and CBF3 datasets, respectively. 
We used subsets of the CBF1 dataset ranging from 2M to 200M series and two CBF3 subsets of 20M and 200M series each. 

The SITS dataset~\cite{pelletier:2019} is a remote sensing dataset (i.e., derived from sensor measurements by satellites orbiting Earth) containing 1M series of size 46 points each, and 24 classes. 
Each series corresponds to 1 pixel of satellite images of the earth, taken at 46 time instances. 
\karima{We drop the last data point for every time series so that we have series of length 45, which can be more efficiently indexed by iSAX2+ with 9 segments of length 5 (remember that all SAX segments should have equal length).}

For ImageNet~\cite{imagenet}, image embeddings were generated using a pre-trained EfficientNetB1~\cite{EfficientNet} neural network. We applied a global average pooling to the last layers of the network to produce a single vector of 1280 real values per image. The dataset contains a total of 1361 distinct classes. For our experiments, we use the vectors of ImageNet's training images ($\approx 1.3$M images) as the series dataset and the vectors of its testing images (50K images from 1000 classes) 
to sample our queries. To also test our methods on a smaller number of classes, we used WordNet's~\cite{wordnet} hierarchical structure and grouped the original classes (``synsets") to 30 larger classes that correspond to the leaf nodes of the hierarchy used by Huang et al.~\cite{Huang2020}.
	
\begin{table}
	\centering
	\caption{Experimental datasets for $k$-NN classification }
	\scalebox{0.88}{
	\begin{tabular}{c|c|c|c|c} 
		Name&Description&Num of series&Length&Classes\\ \hline
		CBF 1 & synthetic & 2M-200M & 128 & 3\\
		CBF 3 & synthetic & 20M / 200M & 128 & 3\\ 
		SITS & satellite images 
		& 1M & 45 & 24 \\ 
		ImageNet & image embeddings & 1.3M & 1280 & 1361 / 30\\ 
	\end{tabular}
	\vspace{-5pt}
	\label{tab:datasets-classification}
	}
\end{table}

\vspace{5pt} 
\noindent{\bf Measures.} We use the following measures to assess the estimation quality of each method and compare their results:

\vspace{5pt} 
\noindent\emph{Coverage Probability:}  It measures the proportion of the time that the prediction intervals contain the true $1$-NN distance. 
If the confidence level of the intervals is $1-\theta$, the coverage probability should be close to $1-\theta$. A low coverage probability is problematic. In contrast, a coverage probability that is higher than its nominal value (i.e., its confidence level) is acceptable but can hurt the intervals' precision. In particular, a very wide interval that always includes the true $1$-NN distance ($100\%$ coverage) can be useless.

\vspace{5pt} 
\noindent\emph{Prediction Intervals Width:} It measures the size of prediction intervals that a method constructs. Tighter intervals are better.  However, this is only true if the coverage probability of the tighter intervals is close to or higher than their nominal confidence level. Note that for progressive distance estimates, we construct one-sided intervals. Their width is defined with respect to the upper distance bound $d_{Q,R}(t)$.

\vspace{5pt} 
\noindent\emph{Root-Mean-Squared Error (RMSE):}  It evaluates the quality of point (rather than interval) estimates by measuring the standard deviation of the true $1$-NN distance values from their expected (mean) values. 

To evaluate the performance of our stopping criteria, we further report on the following measures:

\vspace{5pt} 
\noindent\emph{Time Savings:} Given a load of queries and a stopping criterion, it measures the time saved as a percentage of the total time needed to complete the search without early stopping.

\vspace{5pt} 
\noindent\emph{Exact Answers:} It measures the number of exact answers as a percentage of the total number of queries. 
For $k$-NN classification, we report on \emph{exact classes}, where we assess the percentage of queries for which the progressive class is the final one.

\vspace{5pt} 
\noindent\emph{Accuracy Ratio:} We also measure the ratio of the accuracy of $k$-NN classification with early stopping to the accuracy of exact $k$-NN classification.

\vspace{5pt} 
\noindent{\bf Validation Methodology.} To evaluate the different methods, we use a Monte Carlo cross-validation approach that consists of the following steps. 
For each dataset, we randomly draw two disjoint sets of data series $\mathcal{W}_{pool}$ and $\mathcal{T}_{pool}$ and pre-calculate all distances between the series of these two sets. 
The first set serves as a pool for drawing random sets of witnesses (if applicable), while the second set serves as a pool for randomly drawing training (if applicable) and testing queries. 
At each iteration, we draw $n_w$ witnesses ($n_w=50$, $100$, $200$, or $500$) and/or $n_r$ training queries ($n_r=50$, $100$, or $200$) from $\mathcal{W}_{pool}$ and $\mathcal{T}_{pool}$, respectively. 
We also draw $n_t=200$ testing queries from $\mathcal{T}_{pool}$ such that they do not overlap with the training queries. We train and test the evaluated methods and then repeat the same procedure $N = 100$ times, where each time, we draw a new set of witnesses, training, and testing queries. 
Thus, for each method and condition, our results are based on a total of $N \times n_t = 20K$ measurements. 

\begin{figure}[tb]
\centering
\footnotesize
\includegraphics[width=0.95\linewidth]{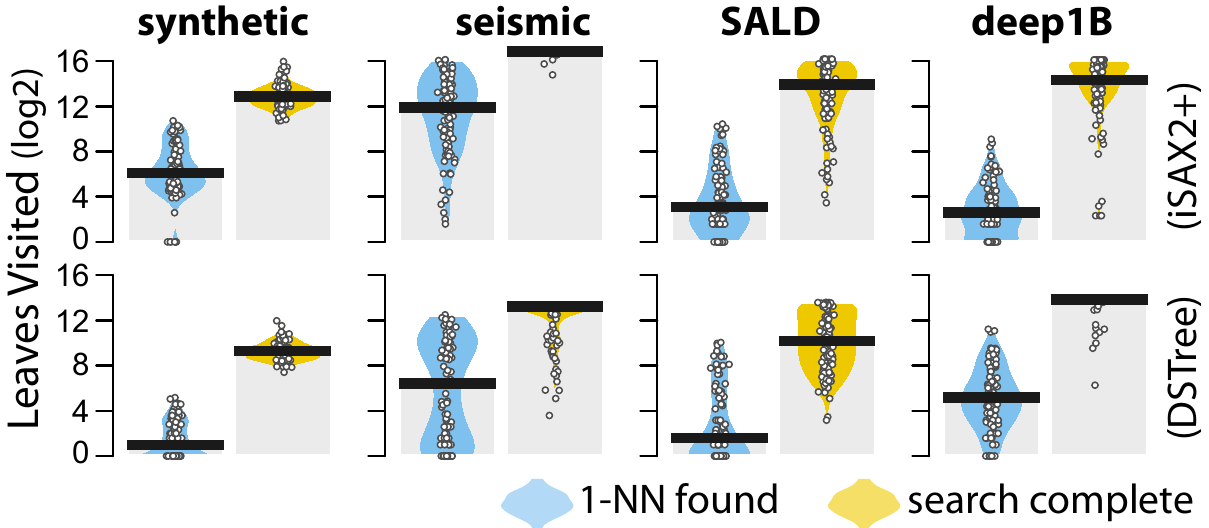}
\vspace*{-2pt}
\caption{Distribution (over 100 queries) of the number of leaves visited (in $log_2$ scale) until finding the $1$-NN (light blue) and completing the search (yellow). The thick black lines represent medians.}
\label{fig:leaves-distributions}
\end{figure}

For all progressive methods, we test the accuracy of their estimates after the similarity search algorithm has visited $1$ ($2^0$), $4$ ($2^2$), $16$ ($2^4$),  $64$ ($2^6$), $256$ ($2^8$), and $1024$ ($2^{10}$) leaves. Figure~\ref{fig:leaves-distributions} shows the distributions of visited leaves for 100 random queries for all four datasets.

\subsection{Results on Prediction Quality}
\noindent{\bf Previous Approaches.} We first evaluate the query-agnostic and query-sensitive approximation methods of Ciaccia et al.~\cite{Ciaccia:1999,Ciaccia:2000}. To assess how the two methods scale with and without sampling, we examine smaller datasets with cardinalities of up to 1M data series (up to 100K for the query-agnostic approach). Those datasets are derived from the initial datasets presented in Table~\ref{tab:datasets} through random sampling. Such smaller dataset sizes allow us to derive the full distribution of distances without sampling errors, while they are sufficient for demonstrating the behavior of the approximation methods as datasets grow. 

\begin{figure}[t]
\centering
\footnotesize
\includegraphics[width=1.\linewidth]{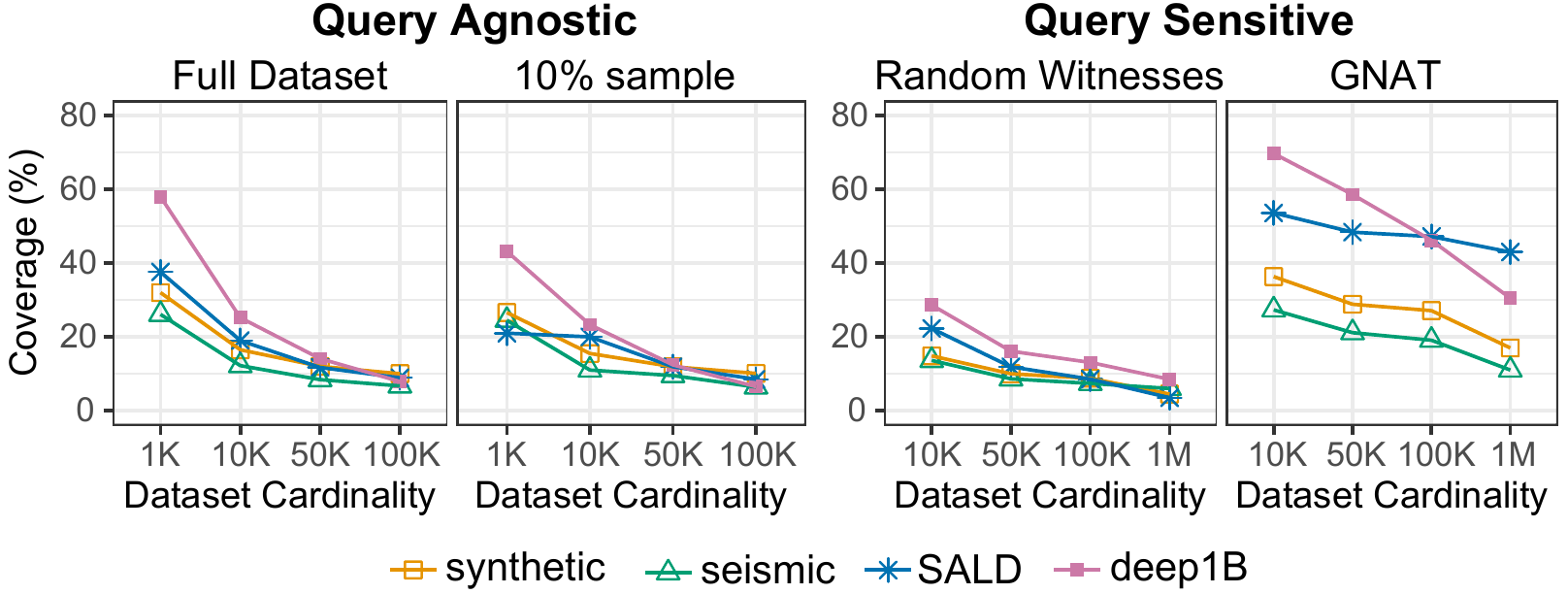}
\vspace{-10pt}
\caption{Coverage probabilities of query-agnostic (left) and query-sensitive (right) 
methods of Ciaccia et al.~\cite{Ciaccia:1999,Ciaccia:2000} for $95\%$ confidence level. We use 500 witnesses for the query-sensitive methods. We show best-case results (with the best $exp$: 3, 5, 12, or adaptive).}
\label{fig:results-ciaccia}
\vspace*{-2pt}
\end{figure}

\begin{figure}[tb]
\centering
\footnotesize
\includegraphics[width=1.\linewidth]{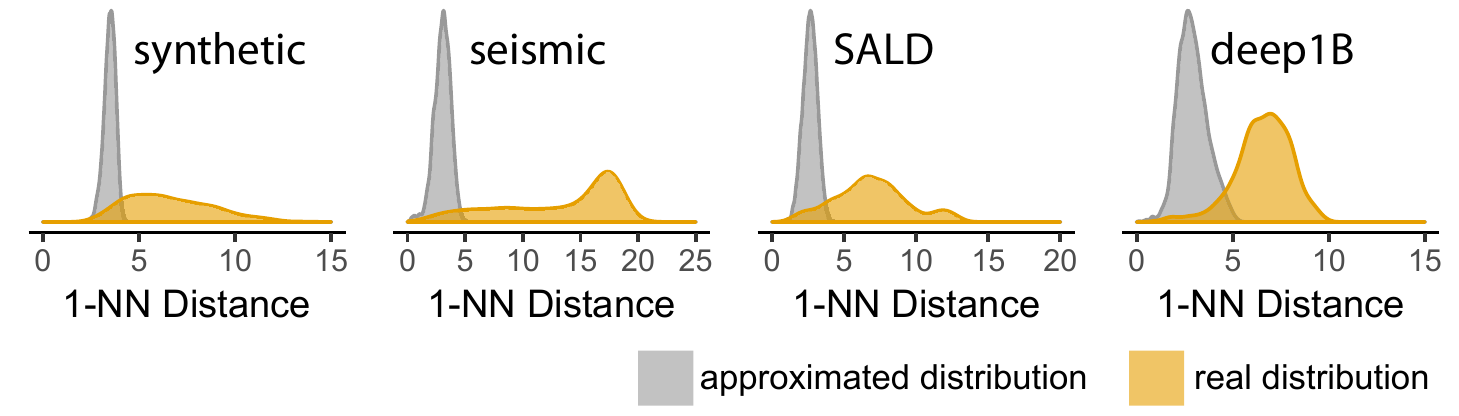}
\vspace{-10pt}
\caption{Real distribution of $1$-NN distances and its query-agnostic approximation based on Ciaccia and Patella~\cite{Ciaccia:2000}. All datasets contain 100K series.}
\label{fig:ciaccia-distributions}
\vspace{-10pt}
\end{figure}

Figure~\ref{fig:results-ciaccia} presents the coverage probabilities of the methods. The behavior of query-agnostic approximation is especially poor. Even when the full dataset is used to derive the distribution of distances, the coverage tends to drop below $10\%$ for larger datasets ($95\%$ confidence level). This demonstrates that the approximated distribution of $1$-NN distances completely fails to capture the real one. Figure~\ref{fig:ciaccia-distributions} compares the real to the approximated distributions for datasets of $100$K series. We observe that the method largely underestimates the $1$-NN distances for all four datasets. 

\begin{figure*}[t]
\centering
\footnotesize
\includegraphics[width=1.0\textwidth]{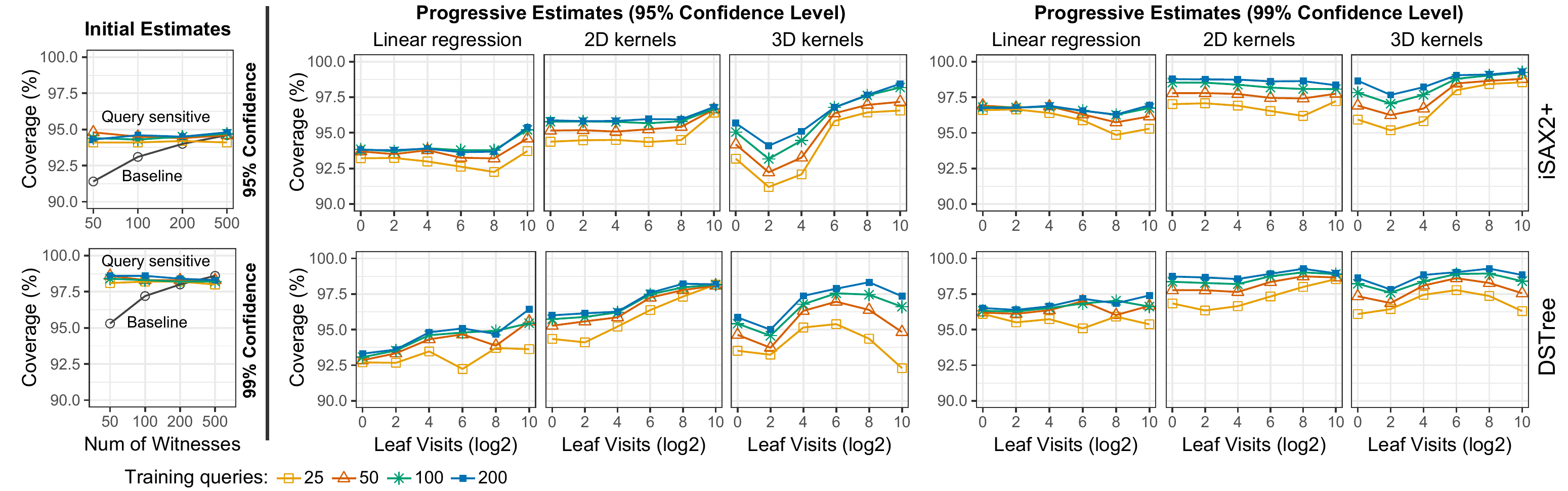}
\vspace{-12pt}
\caption{Coverage probabilities of our estimation methods for $95\%$ and $99\%$ confidence levels. We show averages for the four datasets (synthetic, seismic, SALD, deep1B) and for 25, 50, 100, and 200 training queries.}
\label{fig:results-coverage}
\vspace{-4pt}
\end{figure*} 

\begin{figure}[t]
	\centering
	\footnotesize
	\hspace*{-5pt}
	\includegraphics[width=1.03\linewidth]{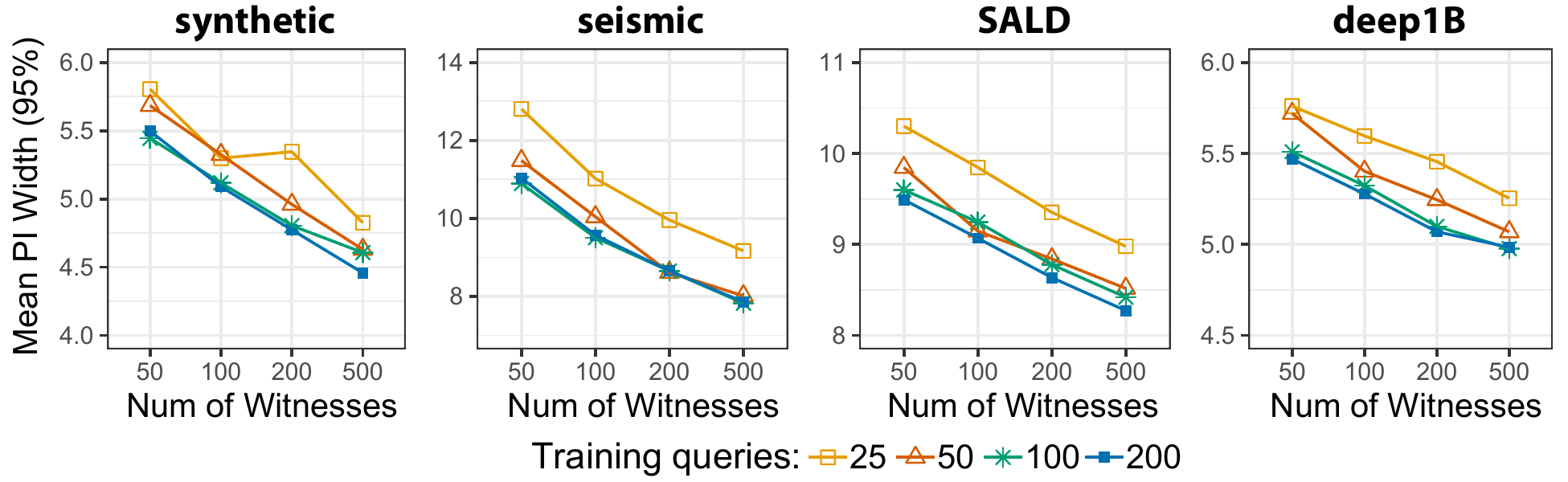}
	\vspace{-12pt}
	\caption{The mean width of the $95\%$ PI for the witness-based query-sensitive method in relation to the number of witnesses and training queries.}
	\label{fig:results-witnesses}
\vspace{-10pt}
\end{figure}

Results for the query-sensitive method are better, but coverage is still below acceptable levels. Figure~\ref{fig:results-ciaccia} presents results for $n_w = 500$ witnesses. Note that our further tests have shown that larger numbers of witnesses result in no or very little improvement, while Ciacca et al.~\cite{Ciaccia:1999} had tested a maximum of $200$ witnesses. To weight distances (see Equation~\ref{eq:weights}), we tested the exponent values $exp = 3$, $5$, and $12$, where the first two were also tested by Ciacca et al.~\cite{Ciaccia:1999}, while we found that the third one gave better results for some datasets. We also tested the authors' adaptive technique.  Figure~\ref{fig:results-ciaccia} presents the best result for each dataset, most often given by the adaptive technique.

We observe that the GNAT method results in clearly higher coverage probabilities than the fully random method. This result is somehow surprising because Ciacca et al.~\cite{Ciaccia:1999} report that the GNAT method tends to become less accurate than the random method in high-dimensional spaces with more than eight dimensions. Even so, the coverage probability of the GNAT method is largely below its nominal level. In all cases, it tends to become less than $50\%$ as the cardinality of the datasets increases beyond 100K, while in some cases, it drops below $20\%$ (synthetic and seismic). 

For much larger datasets (e.g., 100M data series), we expect the accuracy of the above methods to become even worse. We conclude that they are not appropriate for our purposes, thus we do not study them further.

\vspace{4pt} 
\noindent{\bf Quality of Distance Estimates.} 
We evaluate the coverage probability of $1$-NN distance estimation methods for confidence levels $95\%$ ($\theta=.05$) and $99\%$ ($\theta=.01$). Figure~\ref{fig:results-coverage} presents our results. The coverage of the \emph{Baseline} method reaches its nominal confidence level for $n_w = 200$ to $500$ witnesses. In contrast, the \emph{Query-Sensitive} method demonstrates a very good coverage even for small numbers of witnesses ($n_w = 50$) and training queries ($n_r = 25$). However, as Figure~\ref{fig:results-witnesses} shows, more witnesses increase the precision of prediction intervals, i.e., intervals become tighter while they still cover the same proportion of true $1$-NN distances. Larger numbers of training queries also help. 

The coverage probabilities of progressive estimates (Figure~\ref{fig:results-coverage}-Right) are best for the 2D kernel density approach, very close to their nominal levels. Linear regression leads to lower coverage, while the coverage of the 3D kernel density approach is more unstable. We observe that although the accuracy of the models drops in smaller training sets, coverage levels can still be considered as acceptable even if the number of training queries is as low as $n_r = 25$. 

\begin{figure}[tb]
\centering
\footnotesize
\includegraphics[width=1.0\linewidth]{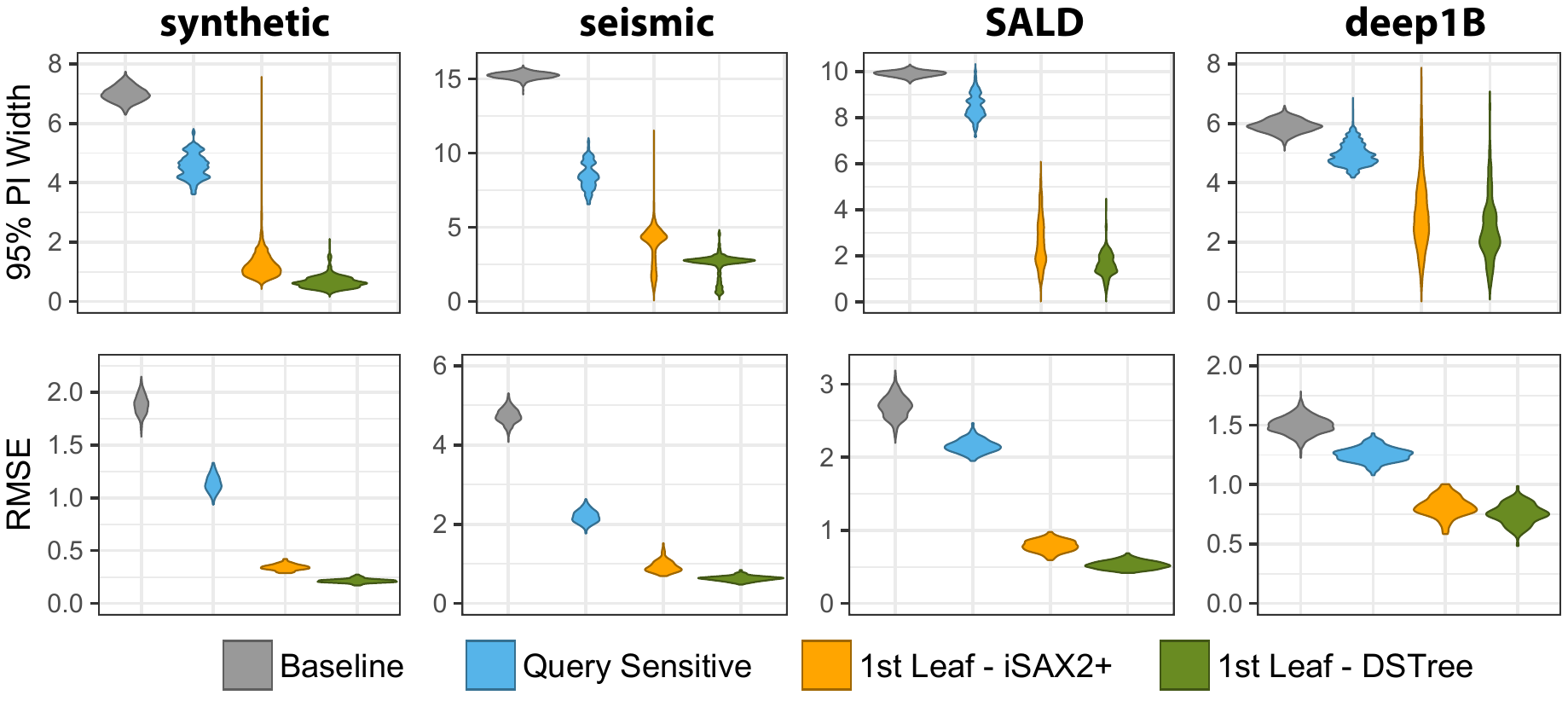}
\vspace{-12pt}
\caption{Violin plots showing the distribution of the width of $95\%$ prediction intervals (top) and the distribution of the RMSE of expected $1$-NN distances (bottom). We use $n_w = 500$ (baseline and query-sensitive method) and $n_r = 100$ (query-sensitive method and 2D kernel model for the 1st approximate answer).}
\label{fig:results-comparison}
\end{figure}

Figure~\ref{fig:results-comparison} compares the quality of initial and early (i.e., based on first approximate answer) estimates provided by different techniques: (i) Baseline, (ii) Query-Sensitive method, (iii) 2D kernel density estimate for iSAX2+, and (iii) 2D kernel density estimate for DSTree.
For all comparisons, we set $n_w = 500$ and $n_r = 100$. For these parameters, the coverage probability of all methods is close to $95\%$. 
We evaluate the width of their $95\%$ prediction intervals and RMSE. We observe similar trends for both measures, where the query-sensitive method outperforms the baseline. 
We also observe that estimation based on the first approximate answer (at the first leaf) leads to radical improvements for all datasets. 
Overall, the DSTree index gives better estimates than iSAX2+.  

\begin{figure}[t]
\centering
\footnotesize
\hspace*{-7pt}
\includegraphics[width=1.05\linewidth]{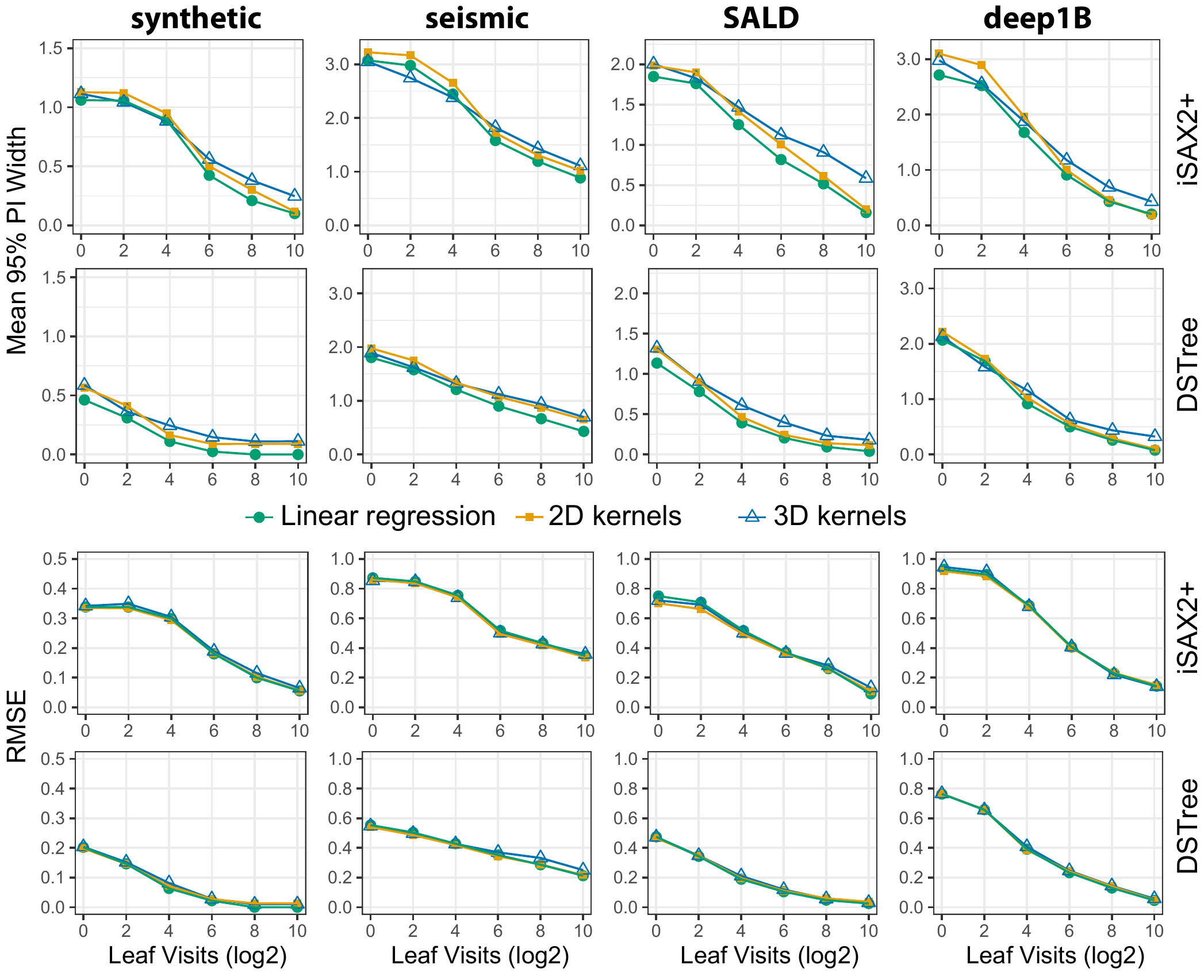}
\vspace{-10pt}
\caption{Progressive models: Mean width of $95\%$ prediction intervals of $1$-NN distance estimates and RMSE. 
Results are based on $n_r=100$ training queries.}
\label{fig:ads-progressive}
\end{figure}

As shown in Figure~\ref{fig:ads-progressive}, progressive answers lead to further improvements. The RMSE is very similar for all three estimation methods, which means that their point estimates are equally good. Linear regression results in the narrowest intervals, which explains the lower coverage probability of this method. Overall, 2D kernel density estimation provides the best balance between coverage and interval width.



\begin{figure}[t]
	\centering
	\footnotesize
	\includegraphics[width=1.0\linewidth]{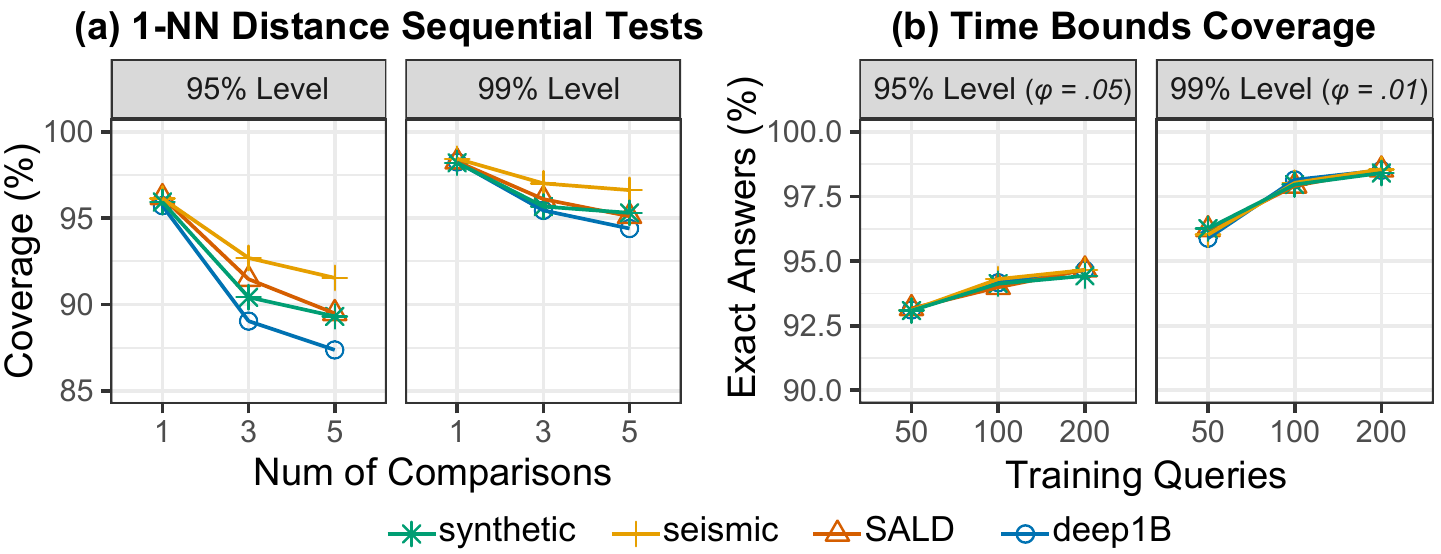}
	\vspace{-12pt}
	\caption{(a) Effect of 3 and 5 sequential tests on the coverage of $95\%$ and $99\%$ prediction intervals. We use 2D kernels with $n_r = 100$. (b) Coverage of exact answers for time upper bounds ($95\%$ and $99\%$ conf. levels).}
	\label{fig:results-sequential}
\end{figure} 

\vspace{4pt} 
\noindent{\bf Sequential Tests.} 
We assess how multiple sequential tests (refer to Section~\ref{sec:progressive}) affect the coverage probability of $1$-NN distance prediction intervals. 
We focus on 2D kernel density estimation ($n_r = 100$), which gives the best coverage (see Figure~\ref{fig:results-coverage}). 
We examine the effect of (i) three sequential tests when visiting 1, 512, and 1024 leaves, and (ii) five sequential tests when visiting 1, 256, 512, 768, and 1014 leaves. 
We count an error if at least one of the three, or five progressive prediction intervals do not include the true $1$-NN distance. 

As results for DSTree and iSAX2+ were very close, we report on their means (see Figure~\ref{fig:results-sequential}(a)). 
The coverage of $95\%$ prediction intervals drops from over $95\%$ to about $90\%$ for five tests (higher for seismic and lower for deep1B). 
Likewise, the coverage of their $99\%$ prediction intervals drops to around $95\%$. 
These results provide rules of thumb on how to correct for multiple sequential tests, e.g., use a $95\%$ level in order to guarantee a $90\%$ coverage in $5$ sequential tests. 
Notice, however, that such rules may depend on the estimation method and the time steps at which comparisons are made. 
An in-depth study of this topic is part of our future work.

\vspace{4pt} 
\noindent{\bf Time Bounds for Exact Answers.}  We are also interested in the quality of time guarantees for exact answers (refer to Section~\ref{sec:exact}). We evaluate the coverage of our time bounds for $50$, $100$, and $200$ training queries for confidence levels $95\%$ ($\phi = .05$) and $99\%$ ($\phi = .01$). 
Figure~\ref{fig:results-sequential}(b) summarizes our results. 
We observe that coverage is good for training samples of $n_r \geq 100$, but drops for $n_r = 50$. 

\begin{figure}[tb]
	\centering
	\footnotesize
	\includegraphics[width=1.0\linewidth]{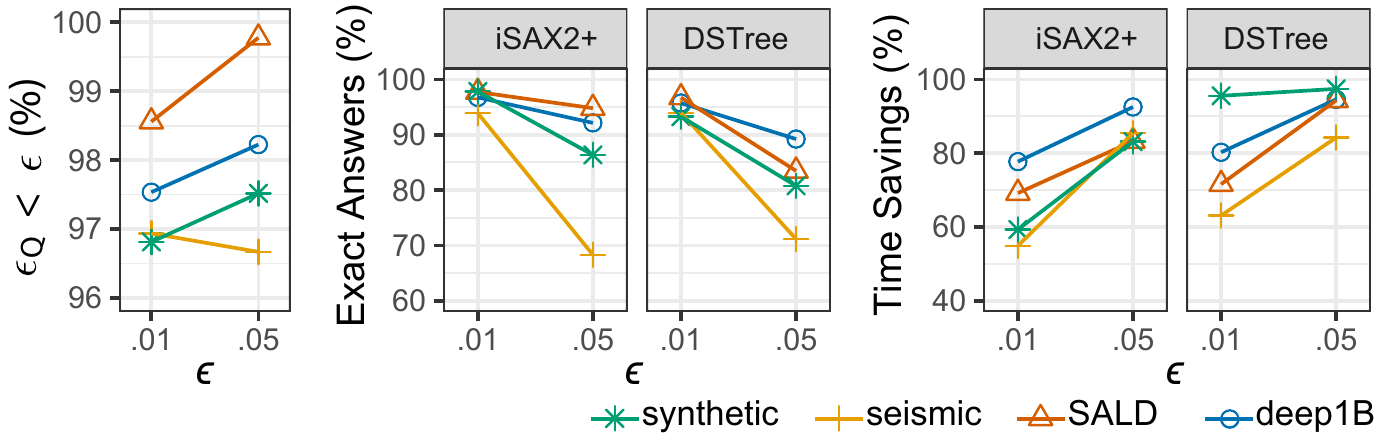}
	\vspace{-12pt}
	\caption{Evaluation of the stopping criterion that bounds the distance error ($\epsilon_Q < \epsilon$). We use $95\%$ prediction intervals ($\theta = .05$) and $n_r=100$ training queries.}
	\label{fig:stopping-criteria-epsilon}
	\vspace{-5pt}
\end{figure} 

\begin{figure*}[tb]
	\centering
	\footnotesize
	\includegraphics[width=1.0\textwidth]{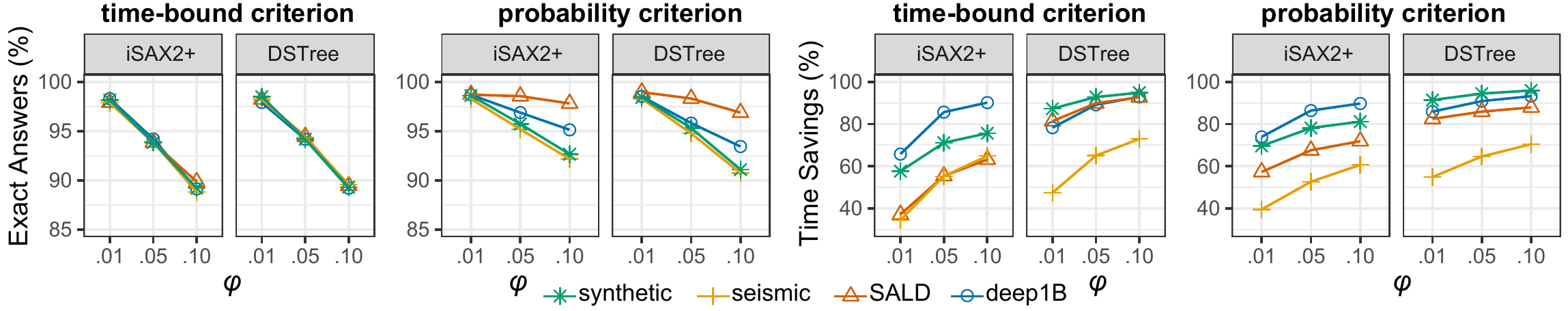}
	\vspace{-10pt}
	\caption{Evaluation of stopping criteria that bound ($\phi$) the probability/ratio of non-exact answers. We measure their ratio of exact answers and their time savings ($\%$). For all conditions, we use $n_r=100$ training queries.}
	\label{fig:stopping-criteria-phi}
\end{figure*} 

\begin{figure*}[tb]
	\centering
	\footnotesize
	\includegraphics[width=1.0\textwidth]{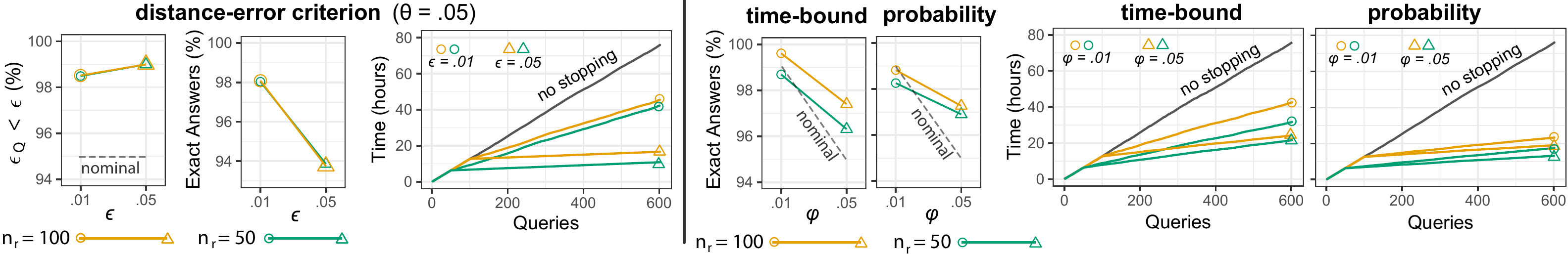}
	\vspace{-10pt}
	\caption{Performance of our stopping criteria for a real workload of 600 queries (deep1B dataset and DSTree). We draw $n_r = 50$ or $100$ random queries for training. We then apply a criterion to the remaining queries. Answers with $\epsilon_Q  < \epsilon$ and exact ones ($\%$) are measured for those ``testing'' queries. (We report means over 100 repetitions.)}
	\label{fig:deep}
\end{figure*}


\subsection{Results on Time Savings}
We compare our stopping criteria (see Section~\ref{sec:stopping}) and assess the time savings they offer. 
Figure~\ref{fig:stopping-criteria-epsilon} shows results for our first criterion that bounds the distance error. 
We consider $16$ discrete and uniform moments $t_i$, where $t_{16}$ is chosen to be equal to the maximum time it takes to find an exact answer in the training sample. 
For each $t_i$, we train an individual 2D kernel density and use $95\%$ prediction intervals ($\theta = .05$) for estimation.   
The coverage (ratio of queries for which $\epsilon_Q  < \epsilon$) exceeds its nominal level ($95\%$) for all datasets, which suggests that results might be conservative. The reason is that stopping could only occur at certain moments. For higher granularity, one can use a larger number of discrete moments. The ratio of exact answers is close to $95\%$ for $\epsilon = .01$ but becomes unstable for $\epsilon = .05$, dropping to as low as $70\%$ for the seismic dataset. 
On the other hand, this results in considerable time savings, especially for DSTree: higher than $90\%$ for the synthetic, SALD, and deep1B datasets.

Figure~\ref{fig:stopping-criteria-phi} compares the two stopping criteria 
that control the ratio of exact answers. 
For the probability criterion, we consider again $16$ discrete moments to stop the search, as above. 
The time-bound criterion results in mean exact answer ratios that are very close to nominal levels, while the probability criterion is rather conservative. 
However, the time gains of the two techniques are comparable. 
For iSAX2+, the probability criterion achieves both a higher accuracy and higher time savings than the probability criterion. 
In contrast, both criteria lead to similar time savings for DSTree, reducing query 
times by up to 95\%.





\vspace{4pt} 
\noindent{\bf Training Costs vs. Gains.} Training linear models with $100$ queries is instantaneous, while learning $16-20$ density functions with 2D kernel density estimation takes no more than 4-6 seconds on a regular laptop. 
Of course, our approach requires the full execution of the training queries. 
For a detailed analysis of the costs of exact similarity search with iSAX2+ and DSTree, we refer the reader to the results of Echihabi et al.~\cite{Echihabi:2018}. Depending on the size and type of the dataset, processing $100$ queries can take some dozens of minutes ($50$ GB datasets), or several hours ($250$ GB datasets). 
Nevertheless, the higher this initial training cost, the higher the benefit is when users later execute their queries. 

Figure~\ref{fig:deep} shows the results for the first 600 queries extracted from a real query workload that comes with the deep1B dataset. 
(Experiment conducted on a server with two Intel Xeon E5-2650 v4 2.2GHz CPUs, 75GB of RAM.)
Results are based on 100 repetitions; each time we draw at random 50, or 100 queries for training. 
We then apply our stopping criteria to accelerate the remaining queries.

The results show that our approach leads to significant performance improvements, while coverage (exact answers, or answers with $\epsilon_Q  < \epsilon$) is very close to, or higher than the nominal levels, even with training sizes of only $50$ queries.
For example, this workload of 600 queries would normally take 76 hours to execute with the DSTree index, but we can execute it in less than 20 hours (probability criterion; including training time), while achieving an average coverage of more than $95\%$ exact answers. 
Finally, we note that, as the trends in the graphs show, the  time savings and speedup offered by our progressive similarity search techniques will increase as the size of the query workload increases. 

\subsection{Results for $k$-NN Similarity Search}
\begin{figure*}[tb]
	\centering
	\footnotesize
	\includegraphics[width=0.99\textwidth]{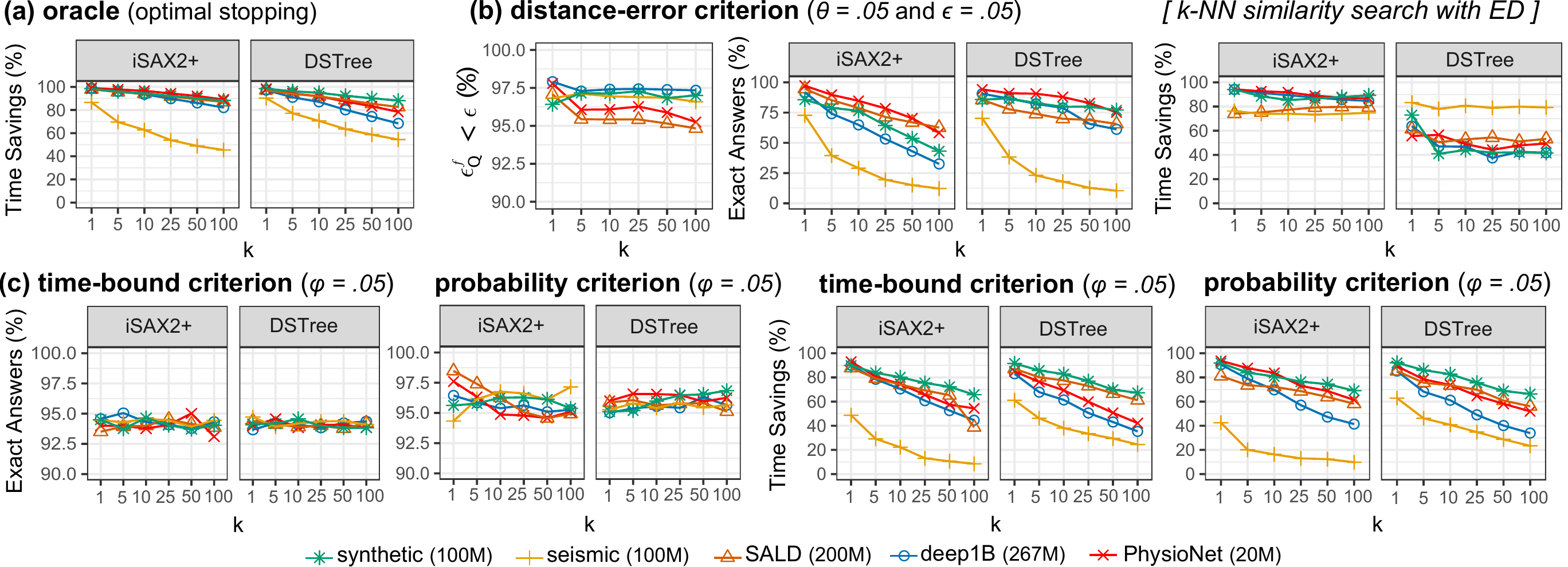}
	\vspace{-6pt}
	\caption{\fanis{Evaluation of our stopping criteria for $k$-NN similarity search ($n_r=100$ training queries). (a) Optimal time savings if an oracle stopped the search as soon as the exact $k$-NN was found. (b) Results for the distance-error criterion, where $\epsilon^f_Q$ refers to the maximum, i.e., family-wise, distance error among all $k$ nearest neighbors. (c) Results for the time-bound criterion and the probability criterion. For our experiments with ED, we used the large versions of the datasets in Table~\ref{tab:datasets}.}}
	\label{fig:knn-search}
	\vspace{-5pt}
\end{figure*}

We evaluate how well our approach generalizes to $k$-NN similarity search. We follow the same experimental method but focus on the performance of our stopping criteria. \fanis{In addition to the four datasets that we used earlier, we also test our approach on the 20M-series PhysioNet dataset (reported in Table~\ref{tab:datasets}).}
Figure~\ref{fig:knn-search} summarizes our evaluation results. 
We first show the average savings in time if stopping was optimally performed by an oracle that knows when a $k$-NN is found (Figure~\ref{fig:knn-search}a). 
We observe that optimal time savings deteriorate as $k$ increases. This is especially the case for the seismic dataset, e.g., savings reduce from $87\%$ ($k=1$) to $45\%$ ($k=100$) when iSAX2+ is used.

For the distance-error criterion, we report results for $\theta = .05$ and $\epsilon = .05$ (Figure~\ref{fig:knn-search}b), where we now control and evaluate the relative family-wise distance error $\epsilon^f_Q$ (see Equation~\ref{eq:family-error}). For iSAX2+, time savings are stable and high (above $70\%$) across the full range of $k$ values. This comes at the cost of a decreasing ratio of exact answers. For DSTree, time savings drop for $k = 5$, but stay constant after, with the exception of the seismic dataset, where time savings stay constant but the ratio of exact answers significantly drops. Overall, the distance-error stopping criterion provides a way for quickly finding low-error answers \fanis{that} may not be exact. 

Figure~\ref{fig:knn-search}c presents results for the time-bound and probability criteria. The probability criterion has a better ratio of exact answers but time savings for both criteria are very similar and drop as $k$ increases. Time savings become as low as $10\%$ for the seismic dataset and $k \geq 50$ (iSAX2+).

\begin{figure*}[tb]
	\centering
	\footnotesize
	\includegraphics[width=0.99\textwidth]{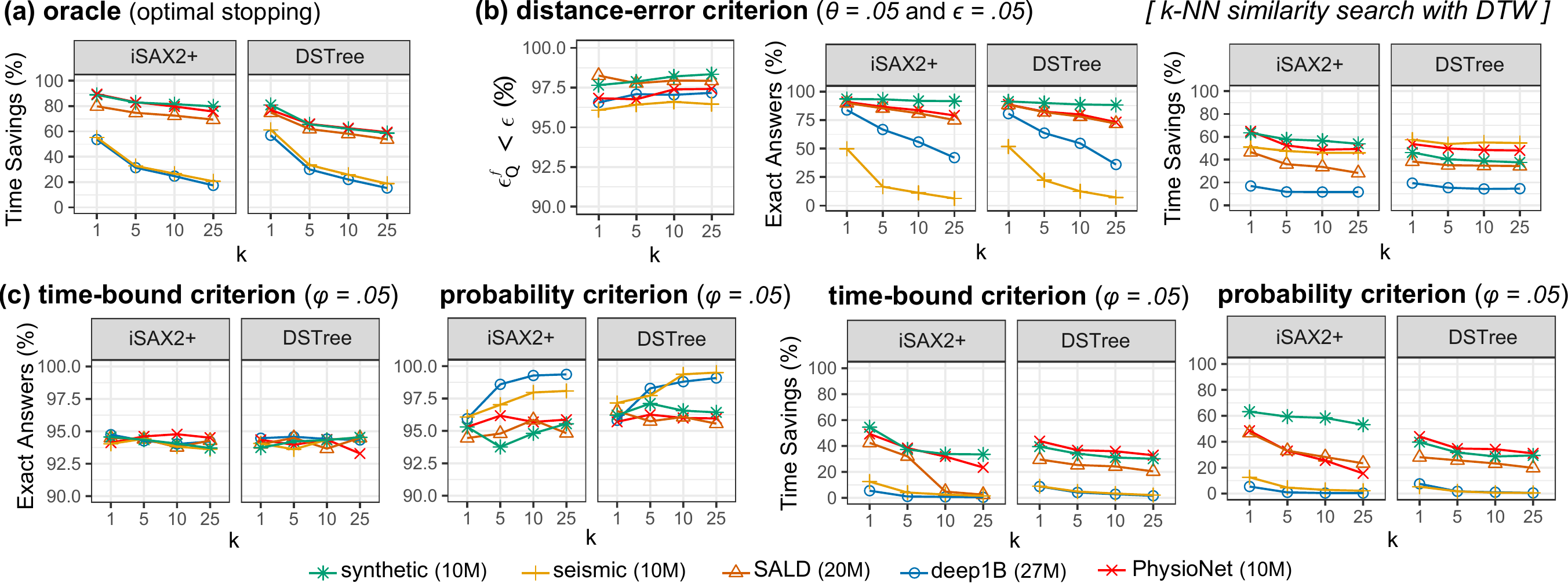}
	\vspace{-6pt}
	\caption{\fanis{Evaluation of our stopping criteria for $k$-NN similarity search and DTW ($n_r=100$ training queries). (a) Optimal time savings if an oracle stopped the search as soon as the exact $k$-NN was found. (b) Results for the distance-error criterion, where $\epsilon^f_Q$ refers to the maximum, i.e., family-wise, distance error among all $k$ nearest neighbors. (c) Results for the time-bound criterion and the probability criterion. For our experiments with DTW, we used the small versions of the datasets in Table~\ref{tab:datasets}.}}
	\label{fig:knn-search-dtw}
	\vspace{-5pt}
\end{figure*}

\fanis{
\subsection{Results for $k$-NN Similarity Search with DTW}
We now repeat the same experiments for DTW, where we apply the bounding envelopes we presented in Section~\ref{section:dtw}.
In this case, we use the smaller versions of the datasets in Table~\ref{tab:datasets}, because similarity search with DTW is extremely expensive for larger datasets. 
For similar performance reasons, we also study similarity search for $k \le 25$. 
We present our results in Figure~\ref{fig:knn-search-dtw}. 

Compared to the results for ED (refer to Figure~\ref{fig:knn-search}), we observe that time savings are now lower for all stopping criteria. 
This loss of performance is due to the wider lower bound distances that we need to use during query answering. 
The exact answer is also found much later. 
This is especially the case for the seismic and deep1B datasets (see Figure~\ref{fig:knn-search-dtw}a). 
For these two datasets, the time-bound and probability stopping criteria result in marginal savings (around $10\%$ for $k=1$), which tend to disappear as $k$ increases (see Figure~\ref{fig:knn-search-dtw}c). 
For the distance-error criterion, which does not require exact answers, time savings are more pronounced. 
For seismic, synthetic, SALD, and PhysioNet, we observe savings in the range of $30-60\%$, independently of $k$ (see Figure~\ref{fig:knn-search-dtw}b). 
Thus, we conclude that our approach is still valuable when using DTW, but with less impressive results, due to the fact that the use of DTW renders the problem harder. 
}

\subsection{Results for $k$-NN Classification}
\label{sec:experiments-classification}

As mentioned earlier, we use a different set of annotated datasets to evaluate our prediction methods for $k$-NN classification. Table~\ref{tab:datasets-groundtruth} presents the $\%$ accuracy of exact $k$-NN classification for the datasets and conditions that we evaluate. It is worth noting that for some datasets, higher accuracy is achieved with a small number of nearest neighbors. 
Experiments on DTW are conducted with smaller synthetic datasets (up to 20M series)\footnote{When using DTW, $k$-NN search becomes computationally very expensive, and the time required to run all experiments with the original, large dataset sizes was prohibitive.}. However, we investigate the role of the dataset size in detail. 
We note also that the accuracy of $k$-NN classification for ImageNet is considerably lower than the accuracy ($> 80\%$) of state-of-the-art neural network architectures~\cite{EfficientNet}. DTW is not meaningful for ImageNet image embeddings while it is very expensive, thus we do not include it in our evaluation.

\begin{table}
	\centering
	\caption{Ground truth ($\%$ accuracy) of the exact $k$-NN classification for the main datasets that we evaluate in our experiments.}
	\scalebox{0.94}{
	\begin{tabular}{c | c | c c c c c} 
		& Dataset & $1$-NN & $3$-NN & $5$-NN & $10$-NN & $20$-NN\\ \hline
		\multirow{5}{*}{\rotatebox[origin=c]{90}{\textbf{Euclidean}}} & CBF1 \scriptsize{(200M)} & 67.0 & 70.2 & 69.7 & 70.7 & 70.8\\
		& CBF3 \scriptsize{(200M)} & 91.0 & 91.2 & 89.8 & 90.8 & 90.8\\
		& SITS & 85.0 & 85.0 & 84.0 & 83.9 & 82.5\\
		& ImageNet & 31.0 & 32.1 & 33.0 & 32.4 & 33.3\\
	        & ImageNet \scriptsize{(30 cl.)} & 57.0 & 58.6 & 58.3 & 56.9 & 56.9\\ \hline \hline
	        	\multirow{3}{*}{\rotatebox[origin=c]{90}{\textbf{DTW}}} & CBF1 \scriptsize{(20M)} & 66.3 & 70.7 & 72.3 & 74.1 & 75.4\\
		& CBF3 \scriptsize{(20M)} & 96.7 & 97.1 & 97.2 & 97.5 & 97.4\\
		& SITS & 84.3 & 83.9 & 83.3 & 82.6 & 81.5\\
	\end{tabular}
	\vspace{-5pt}
	\label{tab:datasets-groundtruth}
	}
\end{table}

\begin{figure*}[tb]
	\centering
	\footnotesize
	\includegraphics[width=0.99\textwidth]{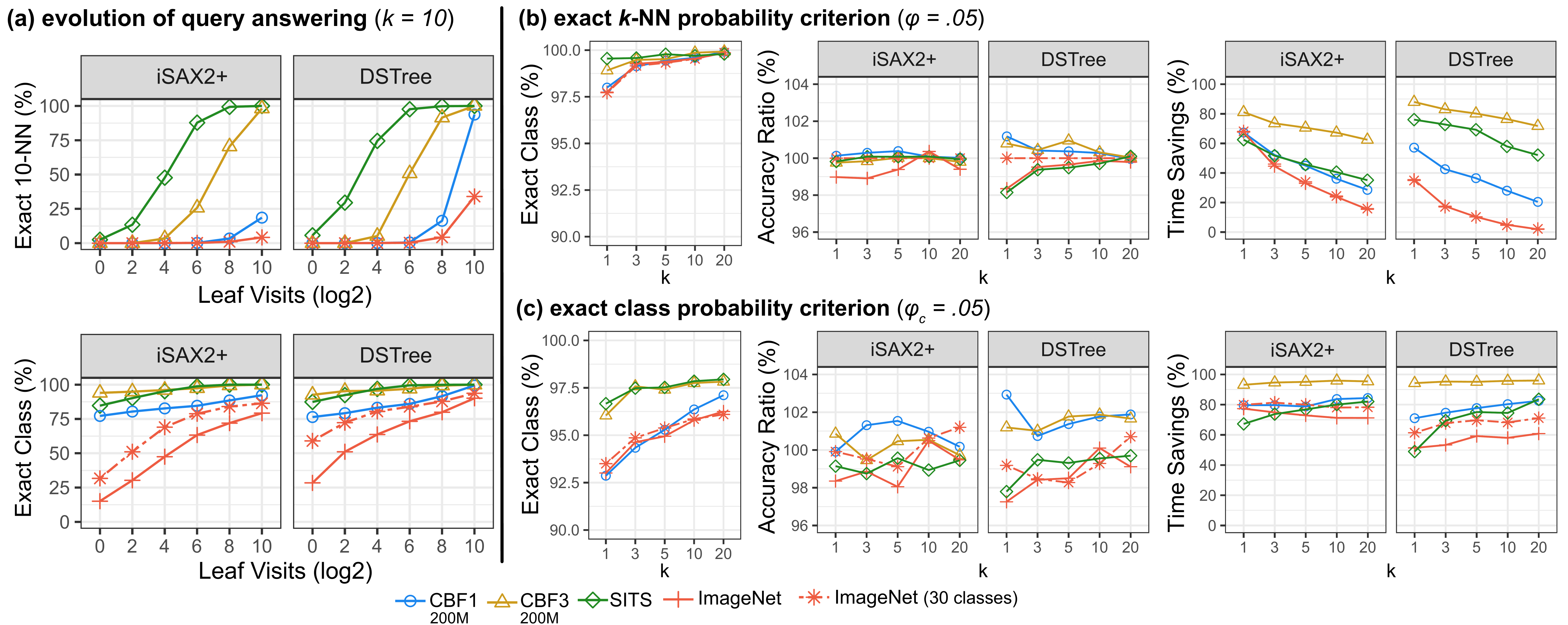}
	\vspace{-2pt}
	\caption{Results for $k$-NN classification (Euclidean distance). (a) Evolution of answers to $10$-NN classification queries for random sample of 1000 queries per dataset. We show the percentage of queries for which the current $10$-NN is exact (top) and the percentage of queries for which the current class is exact (bottom). (b) Results for our conservative stopping criterion that assesses the probability that the current $k$-NN is exact. (c) Results for our class-level criterion that assesses the probability that the current class is exact. We use $n_r = 100$ training queries.}
	\label{fig:knn-classification}
	\vspace{-5pt}
\end{figure*}

We present an overview of how $k$-NN similarity search and classification ($k=10$) evolve in the case of Euclidean distance in Figure~\ref{fig:knn-classification}a. The top graphs show the percentage of queries for which the $10$-NN is found, while the the bottom graphs show the percentage of queries for which the current class is the exact. We observe that for CBF3 and SITS, the exact class is generally found very early, e.g., at the very first leaf. ImageNet turns to be significantly harder -- exact answers arrive much later during the $k$-NN search. In summary, there is no guarantee that a quick approximate answer will return the exact class.

\begin{figure*}[tb]
	\centering
	\footnotesize
	\includegraphics[width=1.00\textwidth]{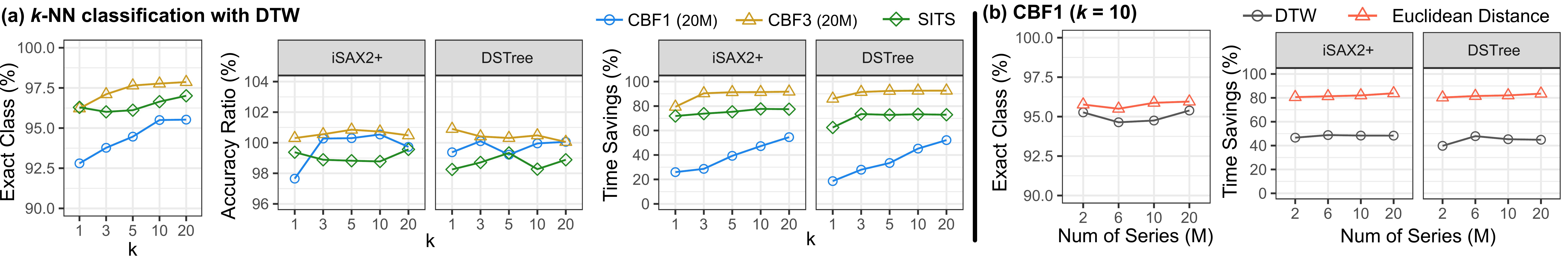}
	\vspace{-18pt}
	\caption{Evaluation of the exact class probability criterion ($\phi_c = .05$) for $k$-NN classification: (a) with DTW and (b) with DTW and Euclidean distance when varying the number of data series for CBF1. We use $n_r = 100$ training queries.}
	\label{fig:knn-DTW}
\end{figure*} 

Figure~\ref{fig:knn-classification}b presents results for the naive probability criterion that controls for the ratio of the exact $k$-NNs ($\phi = .05$). As expected, this criterion is extremely conservative: when $k \ge 3$, the ratio of exact answers becomes higher than $99\%$ for all datasets. As a consequence, time savings drop as $k$ increases and become especially low for the harder datasets. Observe that the accuracy ratio is strictly higher than ratio of exact classes and is often higher than $100\%$. 
Although this result may seem counter-intuitive, we note that an exact answer (i.e., the answer of the non-progressive $k$-NN classifier) is not necessarily correct. In this case, it may happen that the non-exact progressive (approximate) answer is the correct one, and this is more likely to occur when the number of alternative classes is small, as it is the case for the two CBF datasets. Still, one might expect that the correct class would coincide more frequently with the final exact answer. Interestingly, this is not always the case, which suggests that our model predictors can often help to optimally stop, e.g., when consensus among the class of nearest neighbors is high.  

Figure~\ref{fig:knn-classification}c presents results for our exact class probability criterion. The average ratio of exact answers is close to or higher than its nominal level of $95\%$ ($\phi_c = .05$) for most cases but deteriorates when $k \le 3$. The accuracy ratio is again high, ranging between $97\%$ and $103\%$. Time savings are especially high for iSAX2+, greater than $65\%$ and up to $95\%$  for CBF3. Overall, these results demonstrate that our approach can achieve huge time improvement with no or with minimal cost in terms of classification accuracy.

Finally, we evaluate our exact class probability criterion ($\phi_c = .05$) with DTW. Results are presented in Figure~\ref{fig:knn-DTW}a. For CBF3 and SITS, time savings are again at similar levels as with Euclidean distance (see Figure~\ref{fig:knn-classification}c). In contrast, savings are now more modest for CBF1 but they grow as $k$ increases. Of course, the size of CBF1 is smaller now. However, as we see in Figure~\ref{fig:knn-DTW}b, the size of the dataset (i.e., the number of series) does not seem to have any clear effect on savings. It also becomes clear that for this harder dataset, our approach leads to relatively larger benefits with Euclidean distance than with DTW.


%
%

\section{Conclusions}
\label{sec:conclusion}


In this work, we argue that two important research questions are how to provide progressive answers for similarity search queries on very large data series collections, and how to couple these answers with probabilistic quality guarantees. 
Providing progressive answers for data series similarity search queries along with probabilistic quality guarantees is an important research problem.
It eliminates wasted time and reduces user waiting times, in cases where improvement in the final answer is not possible. 

In this context, 
we studied the problems of $k$-NN similarity search and classification for the Euclidean and DTW distance measures. 
We described our approach, ProS, which comprises the first scalable and effective solutions to these problems, and demonstrated its applicability, effectiveness and significant time savings using several synthetic and real datasets from diverse domains. 

As part of our future work, we are going to study in detail how and when such probabilistic measures help humans to effectively complete their visual analysis tasks, as well as which visualization  and human-computer interaction approaches are the most suitable in this context. 
Given the increasing popularity of data series analysis tasks, these research directions are both relevant and important, offering exciting research opportunities. 


\vspace{5pt} 
\noindent
{\bf Acknowledgments.} 
We would like to thank Siddharth Grover for his contributions in the implementation of some of the algorithms in this paper.
Work partially supported by program Investir l'Avenir and Univ. of Paris IDEX Emergence en Recherche ANR-18-IDEX-0001, EU project NESTOR (MSCA {\#}748945), and FMJH Program PGMO with EDF-THALES.

\bibliographystyle{spmpsci}      
\bibliography{ref}

\end{document}